\documentclass[acmsmall]{acmart}

\usepackage{amsmath}
\newtheorem{definition}{Definition}
\usepackage{enumitem}
\usepackage{graphicx}
\usepackage{booktabs, makecell, multirow}
\usepackage{subcaption}
\usepackage[ruled,vlined]{algorithm2e}
\usepackage{float}
\usepackage{bbm}
\usepackage{balance}
\usepackage{makecell}
\usepackage{xcolor}
\usepackage{multirow}
\usepackage{pifont}
\usepackage{color, colortbl}
\usepackage[normalem]{ulem}
\usepackage{appendix}

\definecolor{Gray}{gray}{0.1}

\newcommand{\system}{\textit{W4-Groups}\xspace}
\newcommand\review[1]{\textcolor{black}{#1}}
\newcommand\minorreview[1]{\textcolor{black}{#1}}

\newcommand\hidden[1]{\textcolor{white}{#1}}
\newcommand{\wwww}{\textit{who-what-when-where}\xspace}
\newcommand{\cmark}{\ding{51}}
\newcommand{\xmark}{\textcolor{lightgray}{\ding{55}}}

\AtBeginDocument{%
  \providecommand\BibTeX{{%
    \normalfont B\kern-0.5em{\scshape i\kern-0.25em b}\kern-0.8em\TeX}}}

\setcopyright{acmlicensed}
\acmJournal{PACMHCI}
\acmYear{2024} \acmVolume{8} \acmNumber{CSCW1} \acmArticle{150} \acmMonth{4} \acmPrice{15.00}\acmDOI{10.1145/3637427}

\received{January 2023}
\received[revised]{July 2023}
\received[accepted]{November 2023}

\begin{document}

%

\title[W4-Groups: Modeling the Who, What, When and Where of Groups]{W4-Groups: Modeling the Who, What, When and Where of Group Behavior via Mobility Sensing}

\author{Akanksha Atrey}
\email{aatrey@cs.umass.edu}
\affiliation{%
  \institution{University of Massachusetts Amherst}
  \country{USA}
}

\author{Camellia Zakaria}
\email{camellia.zakaria@utoronto.ca}
\affiliation{%
  \institution{University of Toronto}
  \country{Canada}
}

\author{Rajesh Balan}
\email{rajesh@smu.edu.sg}
\affiliation{%
  \institution{Singapore Management University}
  \country{Singapore}
}

\author{Prashant Shenoy}
\email{shenoy@cs.umass.edu}
\affiliation{%
  \institution{University of Massachusetts Amherst}
  \country{USA}
}

\renewcommand{\shortauthors}{Akanksha Atrey, Camellia Zakaria, Rajesh Balan, \&  Prashant Shenoy} 

\begin{abstract}
Human social interactions occur in group settings of varying sizes and locations, depending on the type of social activity. \review{The ability to distinguish group formations based on their purposes transforms how group detection mechanisms function. Not only should such tools support the effective detection of serendipitous encounters, but they can derive categories of relation types among users. Determining} \emph{who} is involved, \emph{what} activity is performed, and \emph{when} and \emph{where} the activity occurs \review{are critical to understanding group processes in greater depth, including supporting goal-oriented applications (e.g., performance, productivity, and mental health) that require sensing social factors.} In this work, we propose \system \review{that captures the functional perspective of variability and repeatability when automatically constructing short-term and long-term groups via multiple data sources (e.g., WiFi and location check-in data).} We design and implement \system to detect and extract all four group features \wwww from the user's daily mobility patterns. We empirically evaluate the framework using two real-world WiFi datasets and a location check-in dataset, yielding an average of 92\% overall accuracy, 96\% precision, and 94\% recall. Further, we supplement two case studies to demonstrate the application of  \system for next-group activity prediction and analyzing changes in group behavior at a longitudinal scale, exemplifying short-term and long-term occurrences.
\end{abstract}

\ccsdesc[500]{Human-centered computing~Ubiquitous and mobile computing}
\ccsdesc[500]{Human-centered computing~Collaborative and social computing}
\ccsdesc[500]{Applied computing~Sociology}

\keywords{group modeling; social interactions; user mobility; next activity prediction}

\maketitle

\section{Introduction}
Humans are social by nature, and many of our everyday routines involve social interactions with others. For example, activities such as having meals, watching movies, and attending meetings and classes are commonly conducted in groups; with family, friends, and colleagues. Many of these daily routines that involve such group interaction also recur periodically, often involving the same group, yielding four key group interaction characteristics: with \emph{who} is the activity performed, \emph{what} activity is performed, \emph{when} and \emph{where} the interaction occurred. Accounting for these group dynamics -- \wwww (W4) -- for a user reveals many vital aspects of their behavior, including productivity \cite{stogdill1972group}, performance \cite{swain2020leveraging}, and mental and social health \cite{barsade2002ripple, zakaria2019stressmon}.

\review{The rise of the Internet, smartphones, and inexpensive sensing technologies have facilitated the existence of digital nomads, referring to individuals who can work from anywhere \cite{jarrahi2021flexible}. The defining characteristics of ``endless mobility'' and ``location-independent lives'' of digital nomads \cite{lee2019social} have spurred strong interest among researchers to employ various digital means to understand an individual's group behavior, which can occur in many community-based spaces.} For example, social media data has been subjected to social network analyses to produce social graphs of a user, reflecting their social circle and users with whom physical interactions are likely to occur. Similarly, smartphone sensors such as Bluetooth for proximity sensing and GPS for location sensing have been used to approximate face-to-face interaction among groups \cite{lee2013sociophone, brown2014architecture}. While much work focuses on modeling, detection, and understanding group behaviors, these efforts have focused on some but not all of the W4 characteristics in tandem. Prior work assumed that groups are known a priori and have focused on what activities are performed, generally, among students on campus \cite{swain2020leveraging} \review{or leverage on gathering data from serendipitous group encounters \cite{de2015group}. We argue that a generalized group detection framework must accomplish two essential criteria: (1) distinguish short-term and long-term group formations based on their underlying purposes, and (2) support multiple data types to accommodate diverse applications that rely on group sensing mechanisms.}

\review{In this work, we model group interactions of a user by emphasizing all four characteristics, \wwww, of the group dynamics to obtain a holistic view of the user's social interaction. Our approach, grounded by the theoretical understanding of groups in \emph{small group research} \cite{mcgrath1984groups},} utilizes mobile trajectories of users as the basis for inferring W4 group behavior. \emph{Who}, in this case, is a proxy of the device modality producing these mobility trajectories unless the device owner gives explicit permission to be de-anonymized and detected. These mobile trajectories consist of a sequence of locations that users visit and transition from one location to the next. By intersecting the mobile trajectories of users, we can infer common locations \emph{where} they are present at the same time or common paths \emph{when} they move together. These intersections yield insights into when, where, and who and can be augmented with \emph{what} activity is being performed. A key challenge in designing such a technique is to avoid false positives from chance colocations. For example, strangers visiting a coffee shop where a user is present may be falsely identified as a group. The number of such false positives can be significant in public spaces, requiring modeling techniques to distinguish them from true interactions \review{in short-term or long-term groups. The generalization capabilities of \system in capturing W4 characteristics using social science theory across sensing modalities distinguishes it from prior work.}


In modeling, implementing, and evaluating \system, we make the following contributions:
\begin{enumerate} 
    \item We present \system, a group modeling framework, \review{driven by social science perspective to derive four key characteristics distinguishing short-term and long-term group formations based on their underlying purposes. Via any mobility sensing means, the system captures group interactions beyond mere colocation by recursively aggregating users into larger groups based on their mobility rhythms.}
    
    \item We evaluate and validate our approach using multiple mobility datasets and show that \system yields high performance and generalizability in inferring individual and group behaviors even using noisy, coarse-grained WiFi data and incomplete social network check-in data. Through a user study, our method achieves up to 93.49\% accuracy and outperforms various baselines and state-of-the-art techniques. Overall, on three separate datasets, \system yields an average of 92\% overall accuracy, 96\% precision, and 94\% recall.
    
	\item  We show the utility of \system in supporting a next-activity prediction application \review{for short-term group interaction and demonstrate the value of \system in analyzing longitudinal behavioral change as another important application use.}
\end{enumerate}

\review{The rest of the paper is divided as follows. In Section \ref{sec:motivation}, we motivate the need and design rationale for \system. Section \ref{sec:system} discusses the system architecture and modeling approach. We describe \system's prototype implementation in Section \ref{sec:prototype}. Sections \ref{sec:evaluation} and \ref{sec:case-study} are system evaluation and case studies demonstrating the applicability of \system. We discuss the privacy and ethical implications of our work affecting the CSCW community in Section \ref{sec:discussion}. Section \ref{sec:related-work} lists the literature review of group modeling techologies, before concluding the paper in Section \ref{sec:conclusion}.}

\section{Motivating \system}
\label{sec:motivation}
\review{This section motivates our approach to automatically model groups based on the theoretical perspectives in small group research. We describe how our work extends to provide an understanding of the ontological relationships between users and their detected group.}

\vspace{-0.5em}
\subsection{Supporting Group Modeling Applications}
\review{There has been increasing research in the CSCW space where group modeling techniques are applied to gather behavioral data streams of individuals and their groups on topics related to health, engagement, and performance. Alder et al.'s study on how passive personal sensing technologies could help resident physicians with their burnout symptoms revealed the potential for tracking a resident's work deviations from their cohort's norm \cite{adler2022burnout}. Separately, Zakaria et al. reported a significant predictor of stress is the deviations between the student and their group's norm, distinguishing casual groups from workgroups \cite{zakaria2019stressmon} and their social identification to respective groups \cite{zakaria2021detection}. In other applications, Fosh et al. demonstrated how groups often reconfigured themselves during a museum visit, balancing their engagement with exhibits with the group's social dynamic \cite{fosh2016supporting}. Beyond CSCW scholarships are works that uncover group membership in a crowd or public environment to strategize business opportunities \cite{dim2014automatic, fosh2016supporting, shen2018snow}, facilitate surveillance of work dynamics \cite{feese2013sensing}, and pandemic-preparedness \cite{zakaria2022analyzing}.}

\review{These cases exemplify the need to detect formed groups across a wide range of scenarios and sensing modalities, similarly motivating the works by de Freitas et al. to propose a group context framework that allows devices to form groups autonomously and share context \cite{de2015group,de2015using}. However, this framework offers better support for applications intended to increase the opportunities for serendipitous group formation.}

\vspace{-0.5em}
\subsection{Defining a Group}
\label{sec:groupinteractions}
\review{We argue for group modeling techniques to account for critical parameters that reflect real-world formations -- not only effective in modeling serendipitous groups but deriving categories of relation types among users -- driven by social science perspectives.}

\review{By definition, a `Group' is two or more individuals connected by social relationships.} No one description can capture the nuances of what constitutes a group as the relationship can be on a spectrum of properties from primitive forms, such as sharing a family name to more complex factors\review{, such as} the same psychological significance or the skills needed to achieve an agreed-upon goal \cite{forsyth2018group}. \review{Unlike prior work that focuses on opportunistically creating group membership by automatically searching for similar contexts \cite{de2015group}, our work begins by modeling groups grounded on the theoretical understanding of small groups.} We adopt the definition of group as per McGrath et al., where ``a group is an aggregation of two or more people who are to some degree in dynamic interrelation with one another'' \cite{mcgrath1984groups}. 

Here, we define interrelation as the involvement of an individual's routined behaviors and the person around them, and the context of the environment in promoting and terminating these routines. In truth, an individual maintains regularity in daily activities throughout the week, either alone, with one or more persons \cite{monk1990social}. \review{The more one \textbf{repeats} the activity, the higher the likelihood they are routine activities. The higher the \textbf{variability} in these activities, the higher the likelihood they are non-routine activities.} We pay attention to daily routine activities that only occur with one or more persons for our group detection mechanism. In terms of size, a group can be in the range of two (e.g., small collectives) to the thousands (e.g., crowd) \cite{simmel1902number}. Small collective groups typically range from two to seven members \cite{james1953distribution}. Arguably, crowds constitute sets of interlocked small collective groups, albeit gravitating to the smallest size of two \cite{hare1976handbook}. We focus on four characteristics:

\begin{itemize}
	\item \textit{When} an interaction occurs, defined by \textbf{temporal} features such as duration and, accordingly, its recurrence.
 
	\item \textit{Where} an interaction occurs,  defined by \textbf{spatial} features such as the location and, accordingly, the characteristics of a place. 
 
	\item \textit{Who} is involved in the interaction,  defined by \textbf{social} features the users themselves represent, the connections they make with one another, and the frequency of these connections.
 
	\item \textit{What} type of interaction will likely occur, defined by the activity involved, combining and assessing the three attributes above.  
\end{itemize}

To formally capture these characteristics, we define a group interaction as:
\begin{definition}
(Group) A group session $g$ is a co-occurrence of multiple users ($\geq 2$) over a spatiotemporal scale for the same interaction type and defined as
\begin{equation}
    g = \{\mathcal{U},e,d,\mathcal{L},a\}
\end{equation}
where $\mathcal{U}$ represents the set of users in the group, $e$ is entry time of all the group members, $d$ is departure time of all the members, $\mathcal{L} = {\bigcup_{i \in \mathcal{U}} l_{i}}$ represents all the unique locations $\mathcal{U}$ visited, and $a = {\text{mode}_{i \in \mathcal{U}} (a_{i})}$ represents the most common activity being conducted.
\label{def:group}
\end{definition}

\subsection{Design Rationale}
\label{sec:design-rationale}
\begin{figure}[t]
    \centering
    \includegraphics[width=\textwidth]{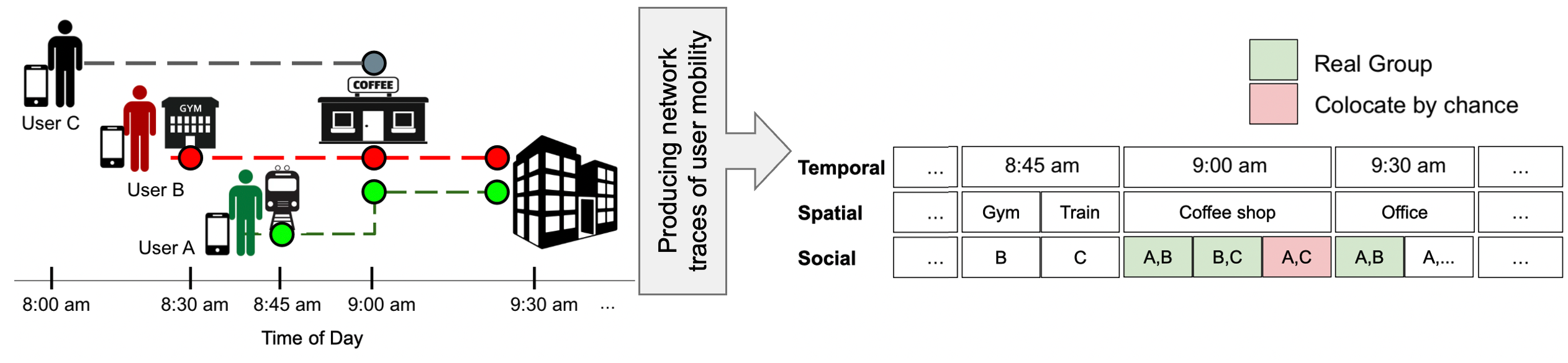}
    \caption{\minorreview{From left to right, network trace of user mobility can produce three key attributes \wwww in the group, reflecting the changes in user location throughout the day. We can infer the pairwise interaction of users collocating by processing these traces collectively at each place over time. Multiple collocations of the same users increase their certainty of existing as a group rather than collocating by chance.}}
    \label{fig:rationale}
\end{figure}

With the prerequisite of a sensing modality, in this case, we assume the user's smartphone with location-based information (i.e., Bluetooth, GPS, WiFi), we can observe the changes in location as they move from one place to the next. Figure \ref{fig:rationale} describes how mobility data of a typical nomadic user can be interpreted at every interval through 24 hours to infer pairwise group interaction with another user. 

As users move or remain stationary in a particular place, their device simply logs the present location. To that degree, the device stands as a proxy of the user. The trajectories of multiple users put together allow us to identify intersecting movements, thereby constructing the following three key attributes; \emph{spatial}, \emph{temporal}, and \emph{social}. The short-lived synchrony between Users A and C and Users B and C at the coffee shop can be interpreted as chance colocation compared to Users A and B, as they would later transition together to an office location and be interpreted as a real group. However, over different time scales of days, weeks, and months, the interpretation of these interactions can change for each pair with repeatable occurrences and their variances over time. The above example yields several critical insights to consider in designing a group detection system. \review{Particularly, an inferred pairwise interaction is more likely to be a real group when their colocations persist across diverse locations (\emph{spatial similarity}), time periods (\emph{temporal similarity}), and involve individuals from the same social circles (\emph{social similarity}). Thus, by capturing the functional perspective of \emph{variability} and \emph{repeatability} when automatically constructing a group, these parameters can help provide a distinctive view of mobility-defined groups and group processes.}

\section{\system System Design}
\label{sec:system}
\review{This section provides an overview of the system's architecture, followed by a detailed description of the design of each component.}

\subsection{System Overview}
\begin{figure}[t]
    \centering
    \includegraphics[width=\textwidth]{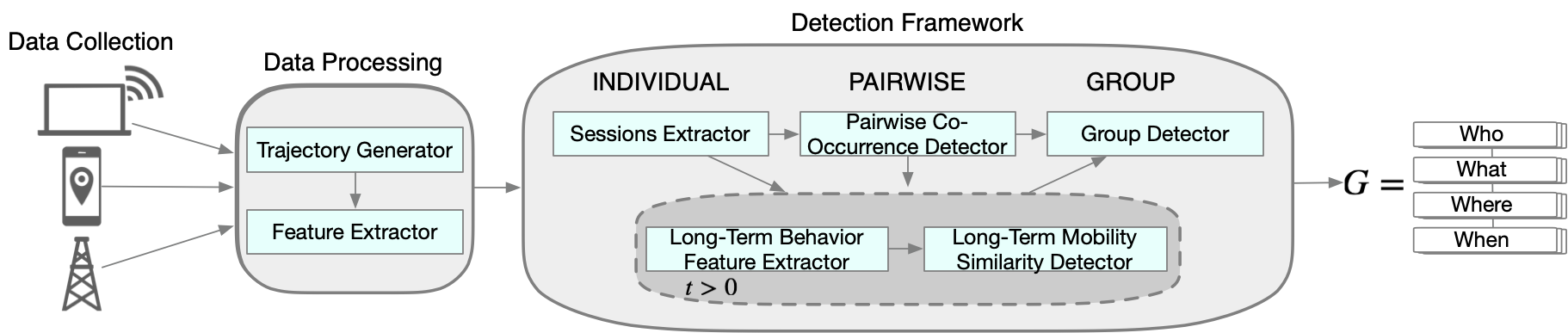}
    \caption{\system overview consisting of two modules: data processing and group detection.}
    \label{fig:pipeline}
\end{figure}

\review{The architectural overview of our system is depicted in Figure \ref{fig:pipeline}. In the design of \system, we overcome several non-trivial challenges associated with mobility traces.}

\begin{enumerate}
    \item \textit{Incomplete and Noisy Data}: Depending on its source, mobility data can range in accuracy and completeness. \review{For example, WiFi traces of connected devices at different access points (AP) are coarse-grained \cite{sen2014grumon}. However, Received Signal Strength Indicator (RSSI) values can detect more precise information \cite{solmaz2020group}. Location-based social networks can be used to gather accurate check-in data but often lack long-term temporal completeness. Thus, our system must consider flexibility in working with varying data granularity.} 
    
    \item \textit{Accurate Detection of Real Groups}: Building on the aforementioned, the type of mobility data can potentially restrict the system's ability to determine the user behind these trajectories. This limitation will result in our system grouping users that may colocate by chance and not for an actual purpose. Our approach must filter out these unintentional groups to accurately detect real groups. 
    
    \item \textit{Computational Complexity}: Since our system looks at every pairwise interaction, determining if a pair belongs to a larger group will increase computational complexity (e.g., O($n^2$) in the simplest case). However, this does not scale for large systems with many users. Our approach reduces the complexity through careful data selection.
\end{enumerate}

\review{The understanding of conceptualizing groups and the four key attributes, described in Section \ref{sec:motivation}, introduced two important factors: \emph{repeatability} and \emph{variability}. Prior works in group detection systems have considered repeated colocations in the time scale of minutes and hours as a measure of real groups \cite{sen2014grumon}. Our system extends this approach to consider long-term dependencies.}

\subsection{Data Processing Module}
\label{sec:data-processing}
Let $U$ be a set of users. We represent user $i$'s mobility trajectories as
\begin{equation}
    M_i = \{\{l_1, t_1\}, \{l_2, t_2\}, ..., \{l_k, t_k\}\}
\label{eq:trajectories}
\end{equation}
where $l$ represents a location (e.g., AP) and $t$ represents a timestamp. These mobility trajectories can be used to extract contextual features such as location type ($b$) and activity ($a$).
Location types, $b_k$, are extracted manually using the buildings' primary purpose. For instance, we can map all campus buildings \review{in WiFi mobility traces} into the following categories: administration, dining, health, labs, landmark, library, parking, police, recreation, residential, school, student organizations, and other. This information can be easily extracted from public sites of colleges, workplaces, and/or cities.

We define activity type as the primary activity performed by the user. For each log event in $M_i$, an activity is extracted as a function of time ($t_k$) and location ($l_k$). We extract six activity types: dining, gym, home, work, transition, and others. \textit{Dining}, \textit{gym}, and \textit{home} activities are extracted using location being a dining location, recreation center, and housing, respectively. \textit{Work} activities are detected using extended periods spent in a particular office building between 8:00 am and 6:00 pm. For a campus setting, we identify facilities that fall into the library, administration, labs, and school type as office buildings. Lastly, \textit{transitions} are detected if the user spends less than ten consecutive minutes at a given location. If the activities do not fall in the remaining five, they are labeled as \textit{other}. This breakdown allows practitioners to associate individual trajectories with tasks and activities and becomes more relevant when dealing with noisy location data such as WiFi signals. Note that the definition of activity is kept generic to protect the user's privacy and can be easily expanded on if explicit permission is received from the user.

\subsection{Group Detection Module}
\label{sec:detection}
\begin{figure}[t]
    \centering
    \includegraphics[width=\columnwidth]{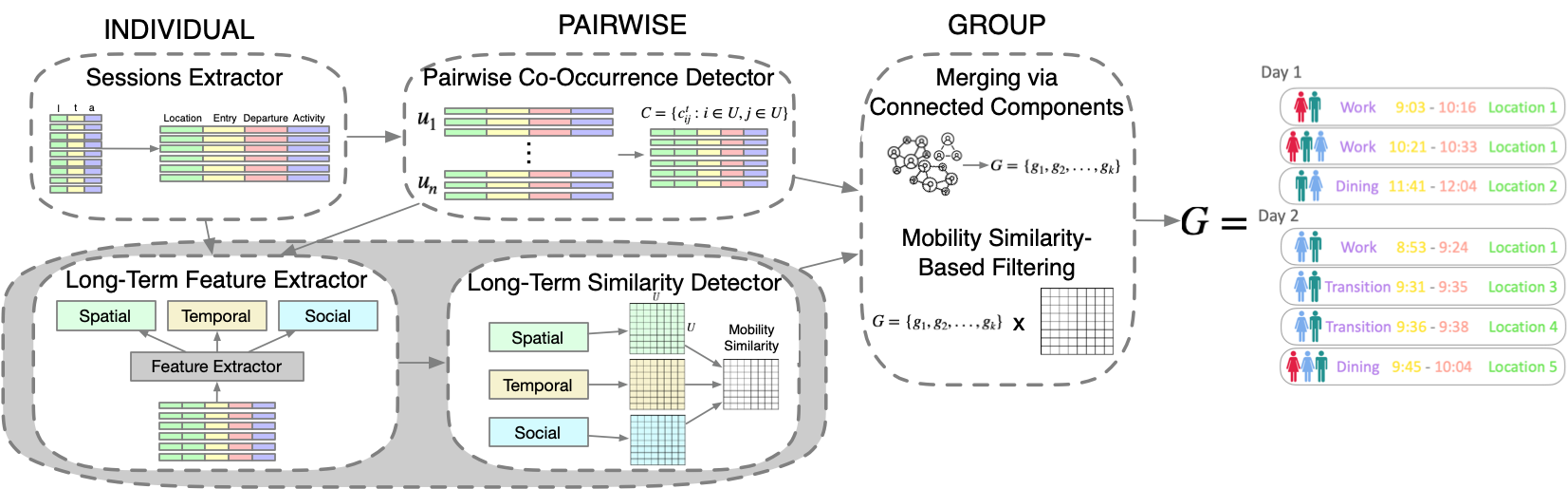}
    \caption{The inner workings of \system's modeling approach to produce \wwww as its output.}
    \label{fig:framework}
\end{figure}

As described in Section \ref{sec:design-rationale}, modeling groups to consider \wwww is primarily based on three primary metrics: \emph{temporal}, \emph{spatial}, and \emph{social similarities}. We now present the modeling approach used by \system's group detection module, depicted in Figure \ref{fig:framework}.

\subsubsection{\review{Modeling Sessions}}
\label{sec:module1}
The first component, sessions extractor, processes features of time and location to output sessions of periods in a vicinity. It aims to minimize errors produced due to poor-quality data by dividing a user's trajectories into periodic sessions. We adopt a strategy proposed by Trivedi et al. to process \review{WiFi-based} mobility trajectories, $M$, by considering all consecutive locations on a single floor of a building (i.e., indoor localization) or all consecutive building locations (i.e., outdoor localization). Note that the degree of information (i.e., room, floor, and building granularity) depends on the localization capability. In doing so, this sub-component outputs \textbf{sessions}, which represent extended periods spent in a particular vicinity \cite{trivedi2020empirical}. 

\begin{definition}
(Session) Given $M_i$, we define each session for user $i$ at time instance $t$ as $s_i^t = \{l,e,d,a\}$ where $l$ represents location, $e$ represents entry time, $d$ represents the departure time and $a$ represents activity type for user $i$.
\end{definition}

A key factor in determining periodic sessions requires temporal information to estimate how long a user spent in a particular location. For incomplete datasets such as check-in data, such information is not available. Thus, when continuous temporal information is unavailable and location information is sparse, each location can be considered a single session. The temporal information can be divided into periods (i.e., 10-minute windows) or time estimates using external means such as location type.

Note that the features in each session are represented as a set since each session can consist of more than one location (i.e., a user who fluctuates between two nearby locations will be classified into a single session). We aggregate these sessions to form a session trajectory over time for user $i$
\begin{equation}
    S_i = \{s_i^1, s_i^2, ..., s_i^k\}
\end{equation}

\subsubsection{Modeling Pairwise Co-Occurrence}
\label{sec:module2}
With each user's session computed, the pairwise co-occurrence detector seeks to discover intersecting trajectories between two users. We define a \textbf{co-occurrence event} $m_{ij}$ between users $i$ and $j$.
\begin{definition}
(Co-occurrence) Given sessions $s_i^t$ and $s_j^t$ of users $i$ and $j$ respectively, if the session locations and times overlap, $m_{ij}^t = 1$ else $m_{ij}^t = 0$.  
\begin{equation}
    m_{ij}^t = 
\begin{cases}
    1, & \text{if } |s_i^t[l] \cap s_j^t[l]| > 0 \text{ and } s_i^t[e] \leq s_j^t[d] \text{ and } s_j^t[e] \leq s_i^t[d]\\
    0, & \text{otherwise}
\end{cases}
\end{equation}
\end{definition}

Based on the positive co-occurrence events, we can extract all \textbf{pairwise co-occurrences} $C=\{c_{ij}^t: {i \in U, j \in U}\}$. Let $c_{ij}^t$ represent a co-occurrence between user $i$ and $j$ at time instance $t$.
\begin{equation}
    c_{ij}^t = \{e, d, l_i, l_j, a\}
\end{equation}

\review{As mentioned earlier, this step presents scalability challenges when dealing with a large number of users.}  Specifically, co-occurrence detection exhibits a $O(n^2)$ complexity where each user's mobility trajectory is compared with every other user. \review{To support a wide range of users, we develop an alternate method to reduce the complexity by carefully \review{selecting pairs for comparison and terminating early} when the overlap between pairs is found. Appendix \ref{appx:fast-pairwise-occurrence} provides a detailed description of this algorithm.}

Note that this component is sufficient in detecting accurate pairwise co-occurrences when dealing with fine-grained location trajectories such as continuous GPS coordinates. However, coarse-grained and incomplete location trajectories will lead to the detection of more false positive pairwise groups. Thus, we employ long-term mobility behavior, described in Sections \ref{sec:module3} and \ref{sec:module4}, to filter out such false positives and understand groups beyond mere co-occurrence.

\subsubsection{Modeling Long-Term Mobility Behavior and Feature Extraction}
\label{sec:module3}
Prior works identify longitudinal similarity between two mobility trajectories using distance measures (e.g., Euclidean distance) \cite{shen2018snow} or time-series correlations \cite{sen2014grumon}. However, these techniques can be inefficient when processing coarse-grained mobility measures like AP connectivity. Moreover, such methods do not allow the inclusion of additional contextual features that are indicative of patterns or routines. Lastly, most existing approaches do not account for the dynamic nature of mobility trends; human interaction and group behavior change and evolve over time. We argue that this is a key factor in identifying generalizable group behavior.

In this work, we define the long-term mobility behavior of individual users and pairs of users in terms of their \emph{spatial}, \emph{temporal}, and \emph{social} attributes, as reasoned in Section \ref{sec:groupinteractions}. This component takes as input user sessions from the sessions extractor component (see Section \ref{sec:module1}) and the pairwise co-occurrence matrix from pairwise co-occurrence detector (see Section \ref{sec:module2}). The output is a set of behavioral factors. All factors are calculated for the period of the long-term data at a user level and pairwise level. We examine different time periods in Section \ref{eval:perf}.

The computation for each group attribute (\emph{spatial}, \emph{temporal}, and \emph{social}) is based on repeatability and variability over the day, as per Table \ref{tab:behavior_features}. For \emph{spatial}, \textit{num\_unique\_locations} is extracted to represent the variability in the locations visited, and \textit{avg\_location\_counts} is extracted to represent the average repeatability in the number of visits to each location. For \emph{temporal}, we extract \textit{num\_unique\_minutes} to represent the variability in the times of the day an individual or pair of users is seen together. We extract \textit{total\_time\_spent} to represent repeatability in time spent (individual-level) and together (pairwise-level). For \emph{social}, \textit{num\_unique\_users} and \textit{num\_interactions} are extracted to represent variability in the number of unique users met and repeatability in the number of interactions across users at the social scale. 

\begin{table}[t]
    \centering
    \caption{Behavioral factors extracted at spatial, temporal, and social scales represent repeatability (Rep) and variability (Var) concepts. Each factor is extracted at the user level and pairwise level.}
    \scalebox{0.8}{
    \begin{tabular}{c|c|l|cc|cc}
        \multirow{2}{*}{\textbf{Feature}} & \multirow{2}{*}{\textbf{Scale}} & \multirow{2}{*}{\textbf{Description}} & \multicolumn{2}{c}{\textbf{Concept}} & \multicolumn{2}{c}{\textbf{Level}} \\
         &  &  & \textbf{Rep} & \textbf{Var} & \textbf{User} & \textbf{Pair} \\
        \hline
        $f^1$ & Spatial & Number of unique locations visited & \xmark & \cmark & \cmark & \cmark \\
        $f^2$ & Spatial & Average number of location visit counts  & \cmark & \xmark & \xmark & \cmark \\
        \hline
        
        $f^3$ & Temporal & Total time spent in minutes & \cmark & \xmark & \cmark & \cmark \\
        $f^4$ & Temporal & Number of unique minutes spent & \xmark & \cmark & \cmark & \cmark\\
        \hline
        
        $f^5$& Social & Number of unique users met & \xmark & \cmark & \cmark & \cmark \\
        $f^6$ & Social & Number of interactions across all users & \cmark & \xmark & \cmark & \cmark
    \end{tabular}}
    \label{tab:behavior_features}
\end{table}

\subsubsection{Long-Term Mobility Similarity Detection}
\label{sec:module4}
In this component, we develop similarity metrics using the features computed in Section \ref{sec:module3} and the Jaccard similarity measure. Jaccard similarity is a statistic used to understand similarity and diversity among sets as follows.
\[J(A,B) = \frac{|A \cap B|}{|A \cup B|}\]

We employ Jaccard similarity as a measure to detect homogeneity among user sessions $S$ at spatial, temporal, and social scales using the features extracted in Section \ref{sec:module3}. Each feature in Table \ref{tab:behavior_features} represents a count of some characteristic in $S$. For spatial features, this characteristic is location $l$. For temporal features, this characteristic is entry and departure times $e$ and $d$. For social features, this characteristic is the relationship with other users in pairwise co-occurrence sessions $C$. 

Let $S^k_i$ and $S^k_j$ represent the sets of $k^{th}$ feature of the mobility trajectories for $i^{th}$ and $j^{th}$ users respectively. The intersection between sessions can be determined using the pairwise-level features such that $|S^k_i \cap S^k_j| = f^k_{ij}$. The union between sets can be calculated using simple set operations between individual and pairwise-level features such that $|S^k_i \cup S^k_j| = f^k_i + f^k_j - f^k_{ij}$ where $f^k_i = |S^k_i|$ and $f^k_j = |S^k_j|$. Thus we define the similarity between users $i$ and $j$ using behavioral features as:

\begin{equation}
    J(S^k_i, S^k_j) = \frac{f^k_{ij}}{f^k_i + f^k_j - f^k_{ij}}
\end{equation}

Accordingly, the definitions of spatial, temporal and social similarities are:

\begin{definition}
(Spatial Similarity) Let spatial similarity $\Phi^{\text{spatial}}_{ij}$ between users $i$ and $j$ represent the homogeneity in spatial features as represented by the number of unique locations visited together versus alone weighted by the frequency of visits. 
\begin{equation}
    \Phi^{\text{spatial}}_{ij} = J(S^1_i, S^1_j) f_{ij}^2
\end{equation}
\end{definition}

\begin{definition}
(Temporal Similarity) Let temporal similarity $\Phi^{\text{temporal}}_{ij}$ between users $i$ and $j$ represent the homogeneity in temporal factors as represented by the total time and number of unique time instances spent together versus alone.
\begin{equation}
    \Phi^{\text{temporal}}_{ij} = \frac{J(S^3_i, S^3_j) + J(S^4_i, S^4_j)}{2}
\end{equation}
\end{definition}

\begin{definition}
(Social Similarity) Let social similarity $\Phi^{\text{social}}_{ij}$ between users $i$ and $j$ represent the homogeneity in social factors as represented by the number of unique common users versus total unique users and number of interactions together versus alone.
\begin{equation}
    \Phi^{\text{social}}_{ij} = \frac{J(C^5_i, C^5_j) + J(C^6_i, C^6_j)}{2}
\end{equation}
where $C_i$, $C_j$ and $C_{ij}$ represent the pairwise co-occurrences including user $i$, $j$ and $ij$ together respectively.
\end{definition}

\review{These metrics capture long-term similarities in \emph{spatial}, \emph{temporal}, and \emph{social} scales, collectively forming a robust measure of long-term mobility behavior as follows.}

\begin{definition}
(Mobility Similarity) Let mobility similarity $\Phi^{\text{mobility}}_{ij}$ between users $i$ and $j$ represent the aggregation of spatial, temporal, and social similarities:
\begin{equation}
    \Phi^{\text{mobility}}_{ij} = \alpha \Phi^{\text{spatial}}_{ij} + \beta \Phi^{\text{temporal}}_{ij} + \gamma \Phi^{\text{social}}_{ij}
\end{equation}
where $\alpha$, $\beta$ and $\gamma$ represent the weight of the spatial similarity, temporal similarity and social similarity respectively such that $\alpha + \beta + \gamma = 1$.
\end{definition}

This definition allows us to weigh each similarity scale based on the input type. For instance, the temporal similarity would be weighted lower in incomplete datasets due to the lack of continuous spatial input. The ability to set $\alpha$, $\beta$, and $\gamma$ makes \system a generalizable group detection system.

\subsubsection{Detecting \system}
\label{sec:module5}
The final component recursively extracts groups using the pairwise co-occurrences from Section \ref{sec:module2} and long-term mobility similarities from Section \ref{sec:module4}. First, the dynamic group detector clusters users connected via pairwise sessions and sharing the same location. That is, if two sets of users $(u_i, u_j)$ and $(u_i, u_k)$ are found at the same location with overlapping sessions, connected components will form the group $(u_i, u_j, u_k)$. To avoid losing information in pairs during recursive merging, all pairs are also included as groups. This outputs all \textbf{groups} $G=\{g_1, g_2, ..., g_k\}$. However, whether groups detected in this step are true groups, or false positive detection is unclear. To filter out such false positives, we identify the group mobility similarity.

\begin{definition}
(Group Mobility Similarity) Let group mobility similarity $\Phi^{\text{group}}_{g_k}$ for all users in group $g_k$ represent the aggregation of $\Phi^{\text{mobility}}$ of all subset of pairs in $g_k[\mathcal{U}]$:
\begin{equation}
    \Phi^{\text{group}}_{g_k} = \text{median}(\Phi^{\text{mobility}}_{ij}  \text{ } \forall \{i \in g_k[\mathcal{U}], j \in g_k[\mathcal{U}]\})
\end{equation}
\end{definition}

Filtering is defined as a function of $\Phi^{\text{group}}_{g_k}$ and group session duration. Particularly, we accept higher duration group sessions ($>60$ minutes), reject low similarity group sessions, and set a higher similarity confidence for smaller duration ($<15$ minutes) for acceptance. This is summarized as:

\begin{equation}
    m_{g_k} = 
\begin{cases}
    1, & \text{if } g_k[d] - g_k[e] > 60 \\
    0, & \text{if } \Phi^{\text{group}}_{g_k} < \phi_l \\
    1, & \text{if } g_k[d] - g_k[e] > 15 \text{ or } \Phi^{\text{group}}_{g_k} > \phi_u   \\ 
    0, & \text{otherwise}
\end{cases}
\label{eq:group_event}
\end{equation}

where $\phi_u$ and $\phi_l$ represent upper and lower bound. The output of \system is all positive group sessions $m_{g_k} = 1$. Although this filtering step requires manual setting of upper and lower similarity bounds, \review{these settings allow for an adaptive system based on different input types.}

\section{Prototype Implementation}
\label{sec:prototype}






We have implemented a complete prototype of \system in Python. Our prototype currently runs on a Dell server with a 2.40GHz Intel Xeon E5-2680 processor and 128GB RAM running Ubuntu 20.04 and can run on a cluster of servers to scale to a large number of users. 


As explained earlier, our approach works with any sensing modality that produces a location and mobility trace of a user. Our current prototype supports two data types: WiFi mobility traces and location-based social network (LBSN) check-in data. \review{A detailed description of the data formats and associated pre-processing steps are provided in Appendix \ref{appx:data-collection}.}

\review{The subcomponents implementation details of \system are as follows.}

\subsection{Sessions Extractor} 
\label{sec:impl_sessions}
The session extraction module computes location-based sessions of each user in two different ways, depending on the dataset. For \textit{temporally continuous datasets} such as WiFi mobility traces where indoor localization is available, we define sessions at floor-level granularity. That is, all consecutive AP associations in the same building and on the same floor are interpolated into the same session. In contrast, for \textit{temporally non-continuous datasets}, such as LBSN check-in mobility traces where localization is at a geo-location level, the module converts each event in trajectory $M$ into 10-minute periods. \review{This discretization can be altered based on the type of LBSN.}

\subsection{Pairwise Co-Occurrence Detector}
In computing trajectory intersections between pairs and their corresponding sessions, the module implements Algorithm \ref{alg:module2} to extract pairwise co-occurrences. In speeding up this computation process, the module uses the Pandas library, allowing the utilization of in-built optimized functions such as \texttt{pd.groupby()} and \texttt{pd.apply()}. Additionally, since users can have more than one device, a common theme with WiFi-based mobility traces, we add a post-processing step that merges overlapping times across devices for pairs.

\subsection{Long-term Mobility Similarity Detector}
\label{sec:parameters}

The following two modules extract long-term behavioral features and determine long-term mobility similarity for each pair of users, respectively. We only gather this long-term information for pairs found in the second component to avoid unnecessary computations.

Our module uses the mobility similarity metric's $\alpha$, $\beta$, and $\gamma$ values in the following ratios:
\begin{enumerate}
	\item \textbf{Equal dependencies:} 0.33-0.33-0.33
	\item \textbf{Dependency on \emph{spatial} and \emph{temporal}}: 0.5-0.5-0.0
	\item \textbf{Dependency on \emph{spatial} and \emph{social}}: 0.5-0.0-0.5
	\item \textbf{Dependency on \emph{temporal} and \emph{social}}: 0.0-0.5-0.5
\end{enumerate}
By default, the $\alpha$, $\beta$, and $\gamma$ values are set to be 0.33, 0.33, and 0.33, respectively. We also compare the length of the long-term data, which by default is set to the full three weeks. Note, this duration is based on our user study data, explained in the following Section \ref{sec:study}.

\subsection{Dynamic Group Detector}
The final module is responsible for extracting W4 groups by recursively merging pair-wise groups and filtering them by their group similarity. In implementing recursive merging, the module expands each session by its location and then groups each group session by session time and location. We employ the \texttt{NetworkX} library for each sub-group found to create a graph and gather connected components. We then work backward to combine each expanded session into groups based on the connected components found. The lower and upper bound similarity values are set to $\phi_l =0.05$ ($75^{th}$ percentile) and $\phi_u = 0.1$ ($95^{th}$ percentile) respectively in Equation \ref{eq:group_event}, unless otherwise stated.

\section{Empirical Evaluation}
\label{sec:evaluation}
In this section, we evaluate the \system prototype. We describe the evaluation setup and present performance, latency, robustness, and generalizability results.

\begin{table}[t]
    \centering
    \caption{\review{Summary of three datasets used to evaluate \system.}}
    \scalebox{0.78}{
    \color{black}{
    \begin{tabular}{l|c|c|c} \hline
    \textbf{} & \emph{\textbf{Campus \#1}} & \emph{\textbf{Campus \#2}} & \emph{\textbf{Gowalla}} \\ \hline

    \textbf{\makecell[tl]{Data Usage}} & \makecell[tc]{System evaluation, Sec. \ref{eval:perf}, \ref{eval:gen}, \ref{eval:latency} \\ Case studies, Sec. \ref{sec:case-study}} & \multicolumn{2}{c}{System generalization, Sec. \ref{eval:gen}} \\ \hline

    \textbf{\makecell[l]{Period/\\Duration}} & \makecell[c]{System eval: Feb-Mar 2022, 3 weeks\\Case Studies: Aug 2019 - May 2022} & \makecell[c]{Aug-Dec 2019,\\1 semester} & \makecell[c]{Jan 2009-Aug 2011,\\32 months} \\ \hline

    \textbf{Location} & \makecell[c]{University of Massachusetts Amherst} & \makecell[c]{Singapore Management University} & \makecell[c]{Worldwide (anonymized)} \\ \hline

    \textbf{No. of Pax.} & 72 (system eval.), 200 (case study) & 57 & 1000 \\ \hline
    
    \textbf{\makecell[tl]{User\\Details}} & \makecell[tl]{
        \emph{Demographics:} 8 M 4 F (deanon. subset)\\
        \emph{Grade:} Graduate (n=11), Staff (n=1) \\
        \emph{Routine:} \makecell[l]{Non-targeted (daily work)}\\
        \emph{Group:} Self-determined teams of 2-4 \\
        \emph{Self reports:} groupwork schedule as \\
        \emph{\hidden{Self reports:}} part of graded material, \\
        \emph{\hidden{Self reports:}} team identification. \\
        \emph{Device MAC Addr:} Smartphones,\\
        \emph{\hidden{Device MAC Addr:}} laptops\\
        \emph{Approval:} Voluntary, IRB-approved \\
        \emph{Compensation:} None \\
    } & 
    \makecell[tl]{
        \emph{Demographics:} 28 M 29 F\\
        \emph{Grade:} Sophomore (n=57)\\
        \emph{Routine:} Targeted (computing course)\\
        \emph{Group:} Pre-assigned teams \\
        \emph{Self reports:} course project schedule \\
        \emph{\hidden{Self reports:}} as graded material, team\\
        \emph{\hidden{Self reports:}} identification assessment \\
        \emph{Device MAC Addr:} Smartphones\\
        \emph{Approval:} Voluntary, IRB-approved \\
        \emph{Compensation:} US$\$$30 \\
    } & 
    \makecell[tc]{N.A.} \\ \hline

    \textbf{\makecell[l]{Ground\\Truth}} & \multicolumn{2}{c|}{\makecell[c]{Individual and group locations\\(datetime,member,activity,relation type)}} & \makecell[c]{Social network\\friendship data} \\ \hline
    
    \textbf{\makecell[l]{Dataset}} & \multicolumn{2}{c|}{.csv files of WiFi traces} & \makecell[c]{.csv files of check-ins} \\ \hline
    
    \end{tabular}}}
    \label{tab:dataset}
\end{table}

\subsection{Study Procedure} 
\label{sec:study}
We employ \review{the following three datasets, with primary analyses using \emph{Campus \#1} dataset. Key details of these datasets are summarized in Table \ref{tab:dataset}}.

\review{We conducted an IRB-approved user study among 72 users, collecting WiFi network traces as mobility data between February and March 2022 from the University of Massachusetts Amherst, labeled as \emph{Campus \#1}\footnote{\review{All WiFi data collection was based on a data usage agreement that includes additional safeguards for anonymization and terms of use. Specifically, device network MAC addresses and usernames used to authenticate the devices to the WiFi network via RADIUS authentication were hashed using strong cryptographic hash functions, anonymizing their identities.}}. A subset (n=12) of students participated in groups, self-determining their teams between 2 to 4 members. They self-reported daily activities of work and social routines,  including when they got out of bed, physically met with their group member/s for that day, start and end time on campus, and meal times. For all the times the group met, participants logged the extent to which each group member was engaged in the activity, from feeling alone to very stimulated. We collected other information, including classes the participant took, their primary work location and device MAC addresses to collect WiFi data.} 

\review{The second dataset, \emph{Campus \#2}, also IRB-approved, is acquired from Singapore Management University, similarly consisting of WiFi network traces as mobility data. Compared to \emph{Campus \#1}, student recruitment was targeted at a computing course cohort, where high groupwork duties were expected as part of the course requirement. The course practices included students pre-assigned to teams of 5 based on their performance in a prerequisite course. While group participation amongst team members was encouraged, opting in/out of the study did not impact class grading and functions. Participants were instructed to have their mobile devices connected to the campus WiFi throughout the semester. They reported their team identification \cite{postmes2013single} with other members on a biweekly basis and supplied their group project schedule, which recorded the time, location, and members present to complete different group tasks. Note participants anonymized the group project schedule, which contained the names of non-consented students.}

\review{Third and finally, we employed a publicly available dataset collected by Yang et al. from \emph{Gowalla} social networking service of check-in information \cite{yang2019revisiting}. It was reported to have 600,000 users in November 2010 and was acquired by Facebook (now Meta) shortly after. This dataset comprises approximately 36 million check-ins made by 300,000 users over 2 million locations. We extract the top 1000 users with the most check-ins for experiments. Ground truth group structure is measured via the users' social network friendship network.}

\review{\review{With WiFi data reflecting users' mobility behavior in datasets \emph{Campus \#1} and \emph{Campus \#2}, the self-reports allow us to verify collocating users' actual relationships and their intentions to be at an area with their group.} Employing \emph{Campus \#2} and \emph{Gowalla} allow us to evaluate model generalizability on datasets collected via varying modalities.}

\subsection{Evaluation Baseline and Metric}
Our evaluation compares \system with eight other methods described below. \review{Several of these methods are chosen as simplified versions of \system where only a component of \system is used standalone for detection. Two of these methods are explicitly selected from prior work that pursued similar goals to our effort.} Specifically, GruMon demonstrated high performance \review{on detecting the \textit{who-when-where} groups} in dense and complex urban spaces such as malls and airports \cite{sen2014grumon}. \review{While Grumon enables location data with additional sensors to enhance group detection, it is one of the few systems that has shown state-of-the-art performance using solely WiFi data (as in our work). We employ Grumon only on raw AP connection events, which are inherently noisy at hour and day-level time scales.} Second, we employ PGT to extract the strength of relationships between pairs by weighting co-location frequency with personal, global, and temporal factors \cite{wang2014pgt}. \review{While PGT is not traditionally used for group detection,} we aim to evaluate how PGT performs in replacement of our long-term similarity detector. Note that these comparisons are drawn based on a re-implementation of both systems.

\begin{enumerate}
    \item \textit{Trajectory Co-Occurrence}: This method includes identifying spatio-temporal co-occurence using raw trajectory data as represented in Equation \ref{eq:trajectories}.
    \item \textit{Session Co-Occurrence}: This method employs only spatio-temporal co-occurrences over sessions. This is equivalent of not employing long-term behavioral data in group detection.
    \item \textit{GruMon* @ Hour-Level}: This method employs an implementation of GruMon, a group monitoring system, at hour-level time windows \cite{sen2014grumon}.
    \item \textit{GruMon* @ Day-Level}: This method employs an implementation of GruMon, a group monitoring system, at day-level time windows \cite{sen2014grumon}.
    \item \textit{Session Co-Occurrence with PGT*}: This method employs session co-occurrences with an implementation of PGT, a method that measures the strength of relationship between pairs.
    \item \textit{Session Co-Occurrence with Spatial Similarity}: This method employs session co-occurrences with only long-term spatial similarity. This is equivalent of setting $\alpha=1$, $\beta=0$ and $\gamma=0$.
    \item \textit{Session Co-Occurrence with Temporal Similarity}: This method employs session co-occurrences with only long-term temporal similarity. This is equivalent of setting $\alpha=0$, $\beta=1$ and $\gamma=0$.
    \item \textit{Session Co-Occurrence with Social Similarity}: This method employs session co-occurrences with only long-term social similarity. This is equivalent of setting $\alpha=0$, $\beta=0$ and $\gamma=1$.
\end{enumerate}

We use accuracy, precision, and recall to measure the effectiveness of our system \review{with the goal that all these metrics should be balanced in a robust system}. 

\subsection{Performance Results}
\label{eval:perf}

\begin{table}[t]
\centering
\caption{Performance of the proposed method against baselines and state-of-the-art methods.}
\scalebox{0.8}{
\begin{tabular}{l|ccc} 
\hline
\textbf{Method} & \textbf{Acc.} & \textbf{Prec.} & \textbf{Rec.} \\
\hline
Trajectory Co-Occurrence & 38.90 & 95.46 & 39.63 \\
Session Co-Occurrence & 61.69 & 61.69 & 100 \\
\hline
Grumon* @ Hour-Level & 33.33 & 96.15 & 32.05 \\
Grumon* @ Day-Level & 72.22 & 85.71 & 80 \\
Session Co-Occurrence with PGT* & 45.18 & 78.21 & 26.99 \\
\hline
Session Co-Occurrence with Spatial Similarity & 82.38 & \textbf{99.15} & 72.05 \\
Session Co-Occurrence with Temporal Similarity & 83.14 & 78.54 & 100 \\
Session Co-Occurrence with Social Similarity & 75.48 & 71.75 & \textbf{99.38} \\
Session Co-Occurrence with Mobility Similarity (\system) &\textbf{93.49} & 97.37 & 91.93 \\
\hline
\end{tabular}
}
\label{tab:performance}
\end{table}

\begin{figure}[t]
    \centering
    \begin{subfigure}[b]{0.32\textwidth}
    \includegraphics[width=\textwidth]{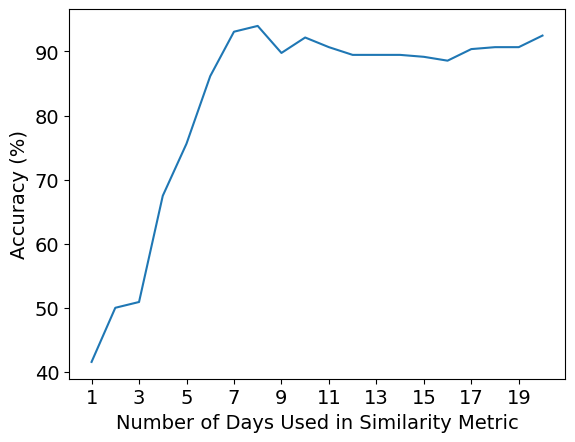}
    \caption{}
    \label{fig:tradeoff_data_len}
    \end{subfigure}
    \begin{subfigure}[b]{0.32\textwidth}
    \includegraphics[width=\textwidth]{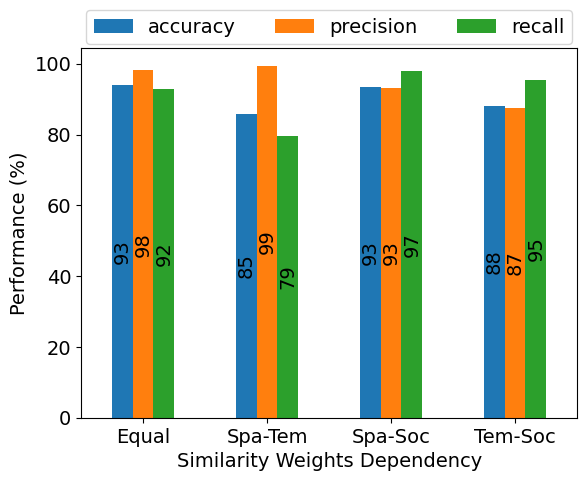}
    \caption{}
    \label{fig:tradeoff_sim_weights}
    \end{subfigure}
    \begin{subfigure}[b]{0.32\textwidth}
    \includegraphics[width=\textwidth]{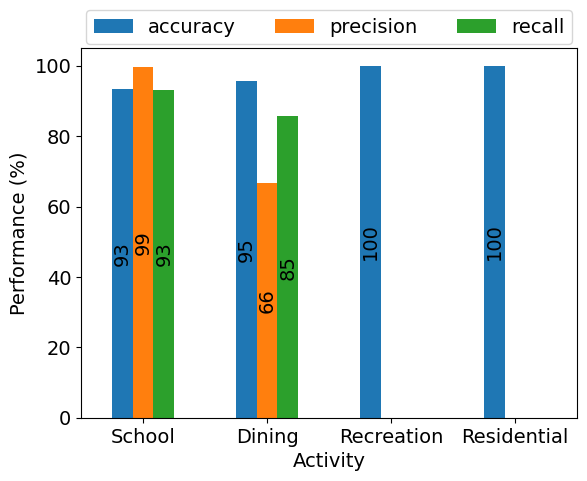}
    \caption{}
    \label{fig:perf_across_activity}
    \end{subfigure}
    \caption{\minorreview{Results of analyzing system parameters and performance tradeoffs: (a) impact of the length of long-term data; (b) impact of the similarity metric weights where 'spa` refers to spatial, 'tem` refers to temporal, and 'soc` refers to social; and (c) performance across group activity types.}}
    \label{fig:solution_exp}
\end{figure}

While \system can be run proactively in a real-time manner as more data arrives, for evaluation purposes, we pass the entire dataset as described in Section \ref{sec:study} to \system. Our first experiment compares \system performance with the baseline methods described above. As tabulated in Table \ref{tab:performance}, \system significantly outperforms all the other methods. Specifically, by including long-term behavior, \system achieved $\approx$ 21\% and $\approx$ 48\% improvements in overall accuracy compared to Grumon and PGT, respectively. While precision and recall are higher for other baseline methods, \system yields the most optimal performance on all three metrics. For instance, although session co-occurrence with spatial similarity has 99.15\% precision, the recall is only 72.05\%, implying only a subset of groups is detected. Similarly, although session co-occurrence with social similarity identifies all correct groups by 99.38\%, it includes many false positives as represented by a 71.75\% precision. Additionally, we note that despite Grumon's high performance at the day level, the output is a coarse cluster of groups instead of fine-grained session-level groups as in \system.

\subsubsection{Tradeoff between length of long-term data and performance} 
Finding the right balance in detection is not straightforward due to the differing nature of groups. While long-term representations are captured through the notion of repeatable colocations and their variances over time, these factors broadly vary across users and groups -- particularly causing errors among idiosyncratic groups when we set general thresholds. 

As illustrated in Figure \ref{fig:tradeoff_data_len}, computing similarity metric based on one day's historical data merely achieves 40\% accuracy across the full dataset. However, it will correctly identify real groups with 90\% accuracy over a week-long history. Performance in detecting real groups tends to stabilize over time, suggesting regularity of users colocating with the same people over the observed duration. This finding implies that while group detection mechanisms can occur in real-time, detecting real groups using coarse-grained data is less likely to be accurate in a short duration; this is where long-term behavioral dependencies prove necessary to proactively update previous results.

\subsubsection{Tradeoff between similarity metric weights and performance}
\review{Figure \ref{fig:tradeoff_sim_weights} charts \ system's performance across different mobility similarity ratios as described in Section \ref{sec:parameters}}. Our results yield $\approx$ 94\% accuracy with equal weights on \textit{spatial}, \textit{temporal}, and \textit{social} attributes to detect groups.

Our observation found most of our student groups colocating in study-related areas and dining halls during these three weeks. However, the frequency of their colocating in these areas also depends on their social relations with their pairs. It should be noted that utilizing \textit{equal dependencies} as weights is optimal for our student population on campus and may not be the same in other settings. In an environment such as a shopping mall, we expect a high distribution of locations grouped by distinct types (e.g., dining, entertainment, groceries, medical, and shopping). In such cases, the weightage on \emph{spatial} similarity may be set more than other factors.

\subsubsection{Performance across activity types} As explained above, during the three weeks of study, the activities of our student groups were highly concentrated in study-related areas and dining halls. For example, the four student groups who went for lunch together were detected by \system with 85.71\% recall. The computation of sessions, as defined in Section \ref{sec:impl_sessions}, is based on consecutive AP associations on the same building floor. Accordingly, what type of activity is being performed is determined by \system at floor-level granularity.

While \system operates to detect groups in residential and recreational locations, as shown in Figure \ref{fig:perf_across_activity}, our results found no positive detection of group activities occurring within recreational and residential typed locations; thus, yielding 0\% on precision and recall. It should be noted that the continuation of indoor masking and social distancing remained as requirements during the study. None of our student groups reported going to the recreational sites.

\begin{figure}[t]
    \centering
    \begin{subfigure}[b]{0.35\textwidth}
    \includegraphics[width=\textwidth]{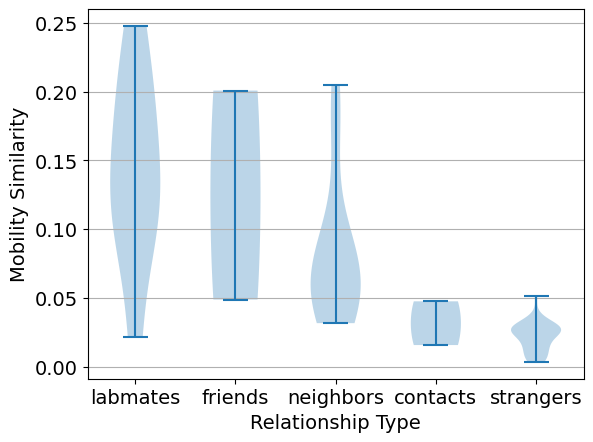}
    \caption{}
    \label{fig:sim_dist}
    \end{subfigure}
    \begin{subfigure}[b]{0.59\textwidth}
    \includegraphics[width=\textwidth]{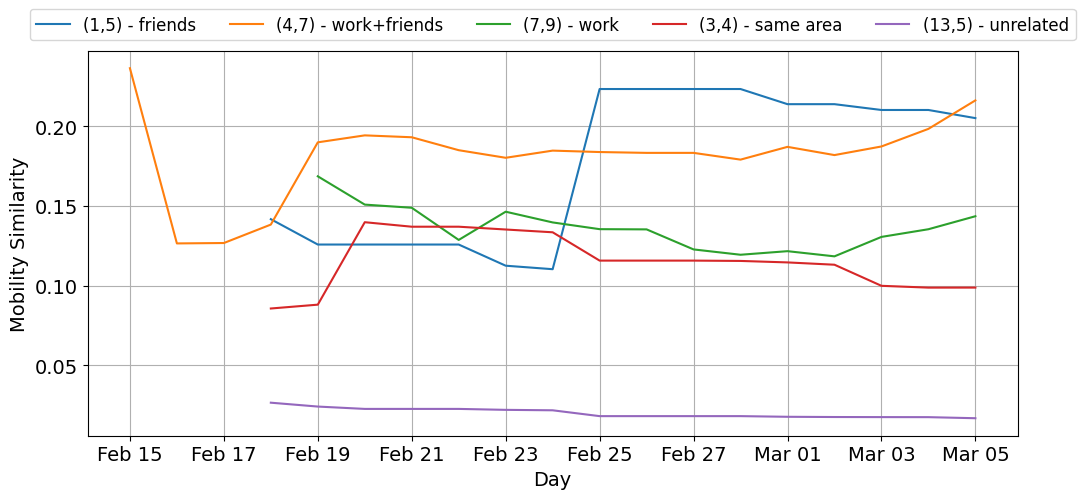}
    \caption{}
    \label{fig:change_in_sim}
    \end{subfigure}
    \caption{\minorreview{Results of analyzing the robustness of long-term behavior across user types: (a) distribution of mobility similarity across pair types; and (b) change in mobility similarity over time for a subset of pairs.}}
    \label{fig:solution_exp}
\end{figure}

\subsubsection{Long-Term Behavior Robustness Across User Types}
\label{eval:robust}
Building on our findings for using long-term data, Figure \ref{fig:sim_dist} plots the probability distribution of student participants for different interactional levels. Among those working in the same lab, there is a higher probability that members in this group take on a mobility similarity value of 0.15 and a lower probability that their mobility similarity value is 0.05. Inversely, two users located on the same floor but do not share the same workspace (e.g., work neighbors) are more likely to have a lower mobility similarity. Through this measure, \system can filter infrequent contacts and strangers who colocate by chance. 

The mobility similarity metric among users who are friends can range from 0.05 to 0.20, likely to be so since there tend to be overlaps in individual perceptions of relationships among members. For example, we learned that five students regarded the interaction with their pair as ``friends'', while three others considered themselves as sharing working and social relationships. Users with overlapping relationships are more likely to share higher similarity metrics as they colocate in the same area for work. We illustrate this example through Figure \ref{fig:change_in_sim}, which plots the changes in the mobility similarity metric of these user groups over time. For groups sharing working and social relationships (i.e., work + friends), their mobility similarity quickly stabilizes near 0.20 over two days. Student groups who regarded themselves as friends met less frequently in work-related areas but more often in dining places. Their mobility similarity, too, stabilizes around 0.20, but it takes approximately a week before \system makes this determination.\newline

\noindent\review{\textbf{Key Takeaway: } The experimental results yield compelling evidence of \system's achieving up to 93.49\% accuracy, even in densely populated areas like dining halls. Including long-term behavioral dependencies contributes to the stability of group detection over time. Furthermore, the analysis reveals the system's ability to accurately differentiate between group relationship types across time. The modular nature of \system enables the customization of similarity metrics to align with specific group settings (e.g., school), further enhancing its flexibility and adaptability.}

\subsection{Model Generalization}
\label{eval:gen}
Our next experiment looks to investigate the generalization of our approach by employing the \emph{Campus \#2} and \emph{Gowalla} datasets. Since \emph{Campus \#2} dataset was gathered as part of a user study that heavily emphasized the analysis of students' interaction during group meetings, we assign less weight to \textit{social} similarity when computing mobility similarity. Particularly, we set $\alpha, \beta, \gamma$ values to be $0.45, 0.45, 0.1$, respectively. We set the lower bound similarity to 0.24 ($75^{th}$ percentile) and upper bound similarity to 0.47 ($95^{th}$ percentile) in Equation \ref{eq:group_event}. For \emph{Gowalla} check-in dataset, we discretize each check-in period into 10 minutes sessions (see Section \ref{sec:impl_sessions}), which inadvertently limits our system's capability to extract temporal information such as the time spent in a particular location. As such, our parameters could only be set for \textit{spatial} and \textit{social} dependencies. We considered tweaking \textit{spatial} similarity with more weight due to its large distribution of locations. Particularly, we set $\alpha, \beta, \gamma$ values to be $0.8, 0, 0.2$, respectively.

\begin{figure}[!t]
\begin{minipage}{0.35\linewidth}
    \centering
    \captionof{table}{Generalized performance across datasets.}
    \scalebox{0.8}{
    \begin{tabular}{l|ccc}\hline
        \textbf{Dataset} & \textbf{Acc.} & \textbf{Prec.} & \textbf{Rec.}  \\
        \hline
        \emph{Campus \#1}: WiFi & 93.49 & 97.37 & 91.93 \\
        \emph{Campus \#2}: WiFi & 93.04 & 100 & 93.04 \\
        \emph{Gowalla} Check-ins & 92.40 & 92.93 & 99.30 \\\hline
    \end{tabular}}
    \vspace{1.5\baselineskip}
    \label{tab:generalizability}
\end{minipage}\hfill
\begin{minipage}{0.54\linewidth}
    \centering
    \captionof{table}{\review{Latency of \system's sub-module by runtime (in seconds) and compute cycles for \emph{Campus \#1}.}}
    \scalebox{0.78}{
    \begin{tabular}{l|cc} 
    \hline
    \textbf{Module} & \textbf{Runtime} & \textbf{\makecell[c] {Cycles}} \\
    \hline
    \makecell[l]{Session Extractor} & 517.16 & 1.240 $\mathrm{e}{+12}$ \\
    \makecell[l]{Pairwise Co-occurrence Extractor} & 17.03 & 4.417 $\mathrm{e}{+10}$  \\
    \makecell[l]{Long-Term Behavior Feature Extraction} & 141.71 & 3.397 $\mathrm{e}{+11}$ \\
    \makecell[l]{Long-term Mobility Similarity Detector} & 0.15 & 3.510 $\mathrm{e}{+8}$ \\
    \makecell[l]{Dynamic Group Detector} & 186.42 & 4.725 $\mathrm{e}{+11}$ \\
    \hline
    \end{tabular}}
    \label{tab:latency}
\end{minipage}
\end{figure}

Table \ref{tab:generalizability} summarizes the performance for detecting groups using \system. Our experiments yield equivalently high performance of 93.04\% accuracy, 100\% precision, and 93.04\% recall for \emph{Campus \#2} dataset, demonstrating that WiFi-based mobility traces, regardless of their setting, can be used to determine groups effectively. Separately, evaluation on the \emph{Gowalla} check-in dataset achieves 92.40\% accuracy, 92.93\% precision, and 99.30\% recall. Despite the incomplete check-in dataset lacking continuous spatial information, the proposed system can detect friend groups accurately using their spatial and social similarities only.\newline

\noindent\review{\textbf{Key Takeaway:} \system's ability to emphasize \textit{spatial}, \textit{temporal}, and \textit{social} similarities according to the nature of interactions contributes to its enhanced generalizability. The results provide quantitative evidence supporting the accurate group detection capabilities of \system across various data granularities.}

\subsection{System Overhead}
\label{eval:latency}
Table \ref{tab:latency} demonstrates the run times of each module in seconds and CPU cycles, respectively, for 72 users over three weeks. The runtime for \system to detect all groups totals up to 862 seconds, assuming each module is executed sequentially. 



By implementing Algorithm \ref{alg:module2}, the pairwise co-occurrence extractor reduces computation time to 17 seconds. As noted, the system overhead is primarily attributed to modules 1, 3, and 5. However, \system parallelizes its operations for modules 1, 2, and 5 in the actual deployment. All three modules operate at a day-wise level, allowing consequent days to be processed through the system concurrently. Modules 3 and 4 require cumulative processing since they work at a long-term level.\newline

\noindent \review{\textbf{Key Takeaway:} \system achieves a total runtime of approximately 14 minutes when processing data from 72 users over three weeks. Due to its parallelizability, modules within \system can be processed concurrently, further reducing the overall runtime.}

\section{Case Studies}
\label{sec:case-study}
In this section, we present two case studies to demonstrate the potential application of \system detection framework. \review{From \emph{Campus \#1}, we meticulously picked 200 anonymous students who frequented a freshmen-designated dormitory. Note the utility of an anonymized dataset meets our computing agreement to only report results at the aggregated level.}

\review{The case studies are motivated by a common theme of analyzing the influence of short-term and long-term groups on individual user behavior. We first model the effect of short-term groups on the mobility of individual users, naturally extending from the \system modeling approach. We then examine the potential of \system in enhancing individual health monitoring systems by analyzing longitudinal changes in group behavior.}


\subsection{Next Activity Prediction}
\label{sec:case-study-1}
Next location prediction has been fundamental in domains ranging from recommendation systems to influencing urban designs. These models use historical spatiotemporal contexts to predict the following location a user will visit. Such predictions can be used in a wide variety of manners. For instance, a mapping service may predict the travel time to the user's next predicted location. 

Prior works have employed Markov and hidden Markov models \cite{gambs2012next, mathew2012predicting}, sequential neural networks \cite{song2016deeptransport, kong2018hst, zhao2020go, feng2020pmf} and more recently, transformer-based models \cite{trivedi2021wifimod} to build human mobility models. However, none of these works have explored predicting beyond the `where' and `when.' One such work by Zhang et al. proposed using hidden Markov models to simultaneously learn mobility models for large group clusters (i.e., who-where-when) \cite{zhang2016gmove}. \review{Conversely, Liao et al. proposed using context-aware recurrent neural networks to predict activity and location for individual users (i.e., what-where-when) \cite{liao2018predicting}.}

This case study \review{examines the influence of groups on individual users in the context of \wwww. Each individual establishes a daily routine consisting of various activities, which can be performed independently or in the company of one or more individuals \cite{monk1990social}. We exploit this aspect of human nature to predict the \wwww of a nomadic user at a fine-grained level using their coarse-grained location trajectories.} We argue that predicting \emph{when} and \emph{where} a user will visit along with \emph{with-who} and \emph{what} activity they will perform is more valuable from urban, campus, and office design perspectives.

\begin{figure}[t]
    \centering
    \begin{subfigure}[b]{0.55\textwidth}
    \includegraphics[width=\textwidth]{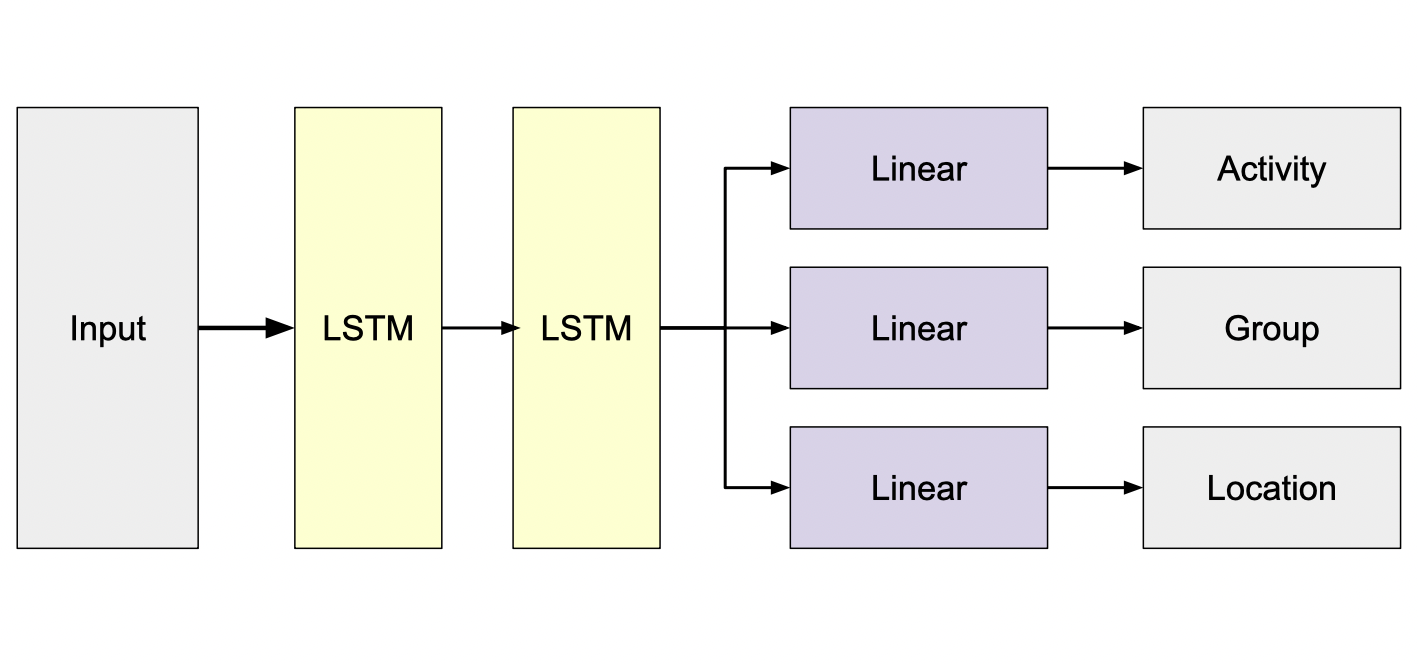}
    \caption{}
    \label{fig:cs1_architecture}
    \end{subfigure}
    \hspace{0.5cm}
    \begin{subfigure}[b]{0.4\textwidth}
    \includegraphics[width=\textwidth]{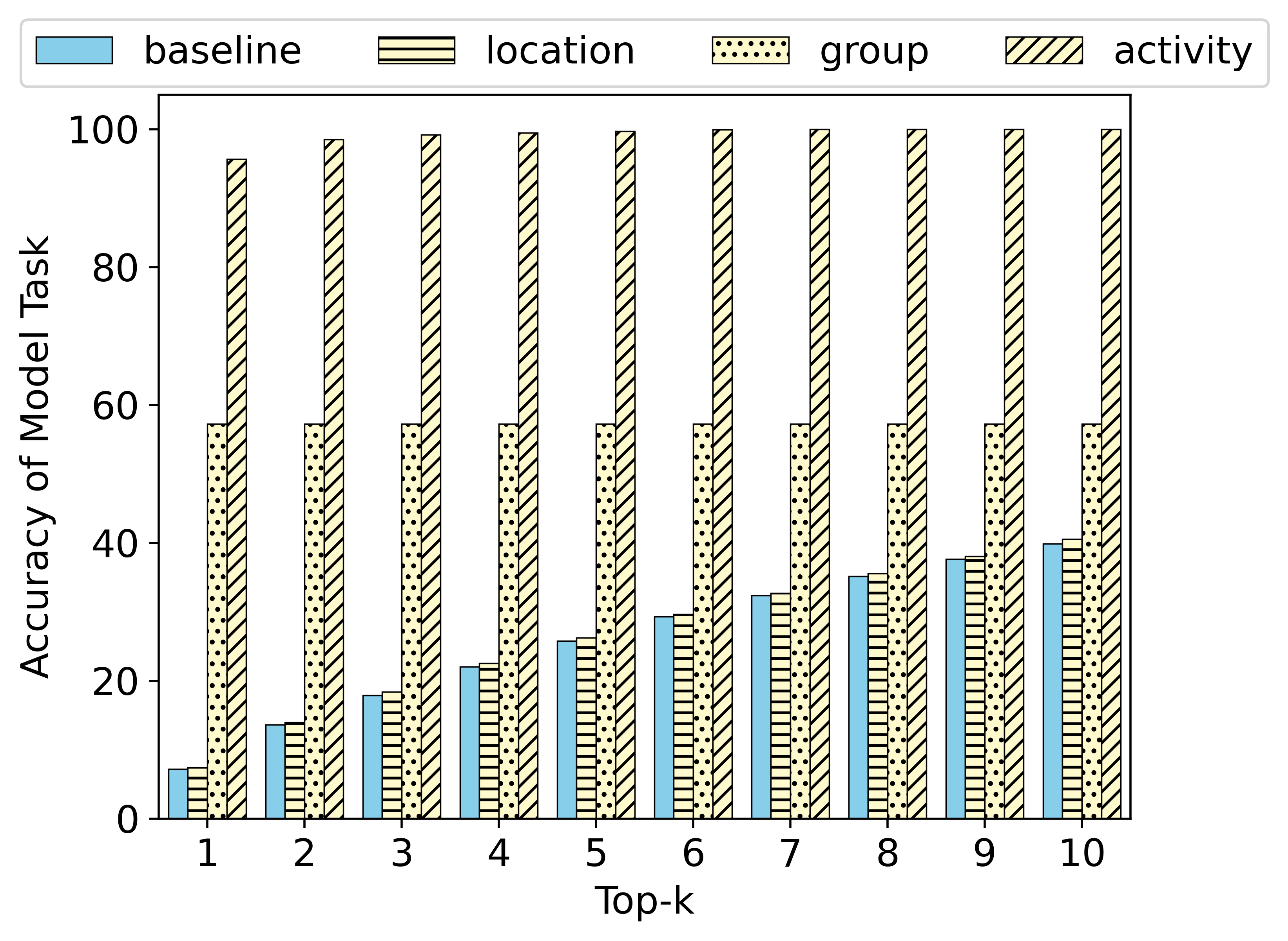}
    \caption{}
    \label{fig:cs1_prediction_bldg}
    \end{subfigure}
    %
    %
    \caption{\minorreview{(a) LSTM architecture for multi-task training; and (b) performance of each task in the model (in yellow) as compared with a traditional next location prediction model (in blue).}}
    \label{fig:solution_exp}
\end{figure}

\subsubsection{Model Training}
We begin by extracting the mobility trajectories of 100 users to train the next activity prediction model. The output of \system is used to annotate individual session trajectories. That is, we augment the session trajectories of each user $i$ to include their group and activity information during time instance $t$ as follows: $s_i^t = \{l,e,d,a,g\}$.

We use this data as input to a long-short term memory (LSTM) model \cite{hochreiter1997long}. \review{LSTMs have shown exemplary performance in predicting human mobility due to their ability to learn and remember long-term dependencies in sequences \cite{kong2018hst, feng2020pmf, trivedi2020empirical, zhao2020go}.}\footnote{\review{While other state-of-the-art models may be employed, such as in Trivedi et al. \cite{trivedi2021wifimod}, LSTMs align with the prevailing approach in the field, allowing for better comparison and benchmarking. It is important to note that our primary goal is to showcase the efficacy of \system for predicting the \textit{who-what-when-where} of a nomadic user, and the specific model choice does not affect this objective.}}
Particularly, we train a two-layer multi-task LSTM model $M$ where input is the previous five session sequences. Output is the next location, next group, and next activity: $M: s_i^{t-5},s_i^{t-4},..,s_i^{t-1} \to l_i^t, g_i^t, a_i^t$\footnote{The choice of architecture is selected based on existing state-of-the-art models for next location prediction.}. The model architecture is depicted in Figure \ref{fig:cs1_architecture}. The distinct linear layers allow for simultaneous training of the three learning tasks by fine-tuning the representations learned by the LSTM layers. For each task, we compute its unique cross-entropy loss and aggregate it for learning. The model is implemented using PyTorch and is trained using a learning rate of $1\mathrm{e}{-4}$ with a weight decay of $1\mathrm{e}{-6}$ and a hidden layer size of $128$. We use batches of size $128$ with a dropout rate of $0.2$ between the LSTM layers and train for $15$ epochs.

\subsubsection{Model Performance}
Figure \ref{fig:cs1_prediction_bldg} presents the results of next activity prediction at building-level localization. We compare the model's performance with a traditional next-location prediction model where the output modality is only `where.` The models are evaluated using the top-k accuracy metric, which counts the instances where the correct label is a subset of the model's top-k most probable classes. There is no notion of top-k for group prediction since it is a multi-class problem.

Location prediction in both models has similar performance; for $k=10$, the models can accurately predict the next location with approximately 40\% accuracy. However, with the multi-output model, we can also identify the next group the user will be with 58\% accuracy and the next activity the user will perform with 99\% accuracy. Further optimization via advanced LSTM cells that contain task-specific parameters for task-specific learning can improve the model's performance \cite{lu2019sc}.\newline

\noindent \textbf{Key Takeaway:} The results demonstrate the applicability of \system in predicting \emph{who-what-where} for the next timestep (\emph{when}). \review{Employing \system to augment traditional next location prediction models with group-level information offers unique capabilities that can enhance our understanding of the underlying reasons behind location visits. Since groups are integral to daily routines, this capability holds significant implications for future research and advancements in various fields, including location-aware mobile services and recommendation systems.}
 


\subsection{Longitudinal Change in Behavior}
\label{sec:case-study-2}

\review{The subject of health and well-being is a multifaceted outcome resulting from the interplay of individual, social, and environmental factors.} Much work in ubiquitous computing has focused on developing health risk management systems to help the population cope with behavioral, mental, and social health-related issues. \review{These systems enable accurate monitoring of psychological measures and estimation of mental states via various modalities, such as smartphone sensing \cite{wang2014studentlife, saeb2015mobile, canzian2015trajectories} and passive WiFi sensing \cite{ware2018large, zakaria2019stressmon, mammen2021wisleep}.} However, the emphasis on sensing individual factors overlooks social interaction that significantly influences one's health. \review{Jetten et al. present compelling evidence emphasizing the importance of social groups in preserving health and well-being, yet they assert that group identities are often disregarded in practical applications \cite{jetten2014groups}.} We believe \system is highly applicable to the extension of such systems, as it has the potential to provide accurate measures related to group behaviors and their social routines.


One aspect of supporting health monitoring systems is the ability to 
\review{understand how user behavior changes over time.} This case study demonstrates how \system \review{supports this goal by effectively monitoring longitudinal changes in group behavior. By integrating group information, our study highlights the potential of enhancing health monitoring systems in capturing and analyzing shifts in individual and collective behavior patterns.}

The WiFi-derived mobile trajectories were passed through the \system pipeline. For pairwise groups, we aggregate their behaviors at per week granularity. We report the total time spent per week in groups across the semester. Since group sessions depend on WiFi connectivity, these sessions may not precisely represent users spending long hours in one place but relatively shorter sessions of users in the vicinity.

\begin{figure}[t]
    \centering
    \begin{subfigure}[b]{0.35\textwidth}
    \includegraphics[width=\textwidth]{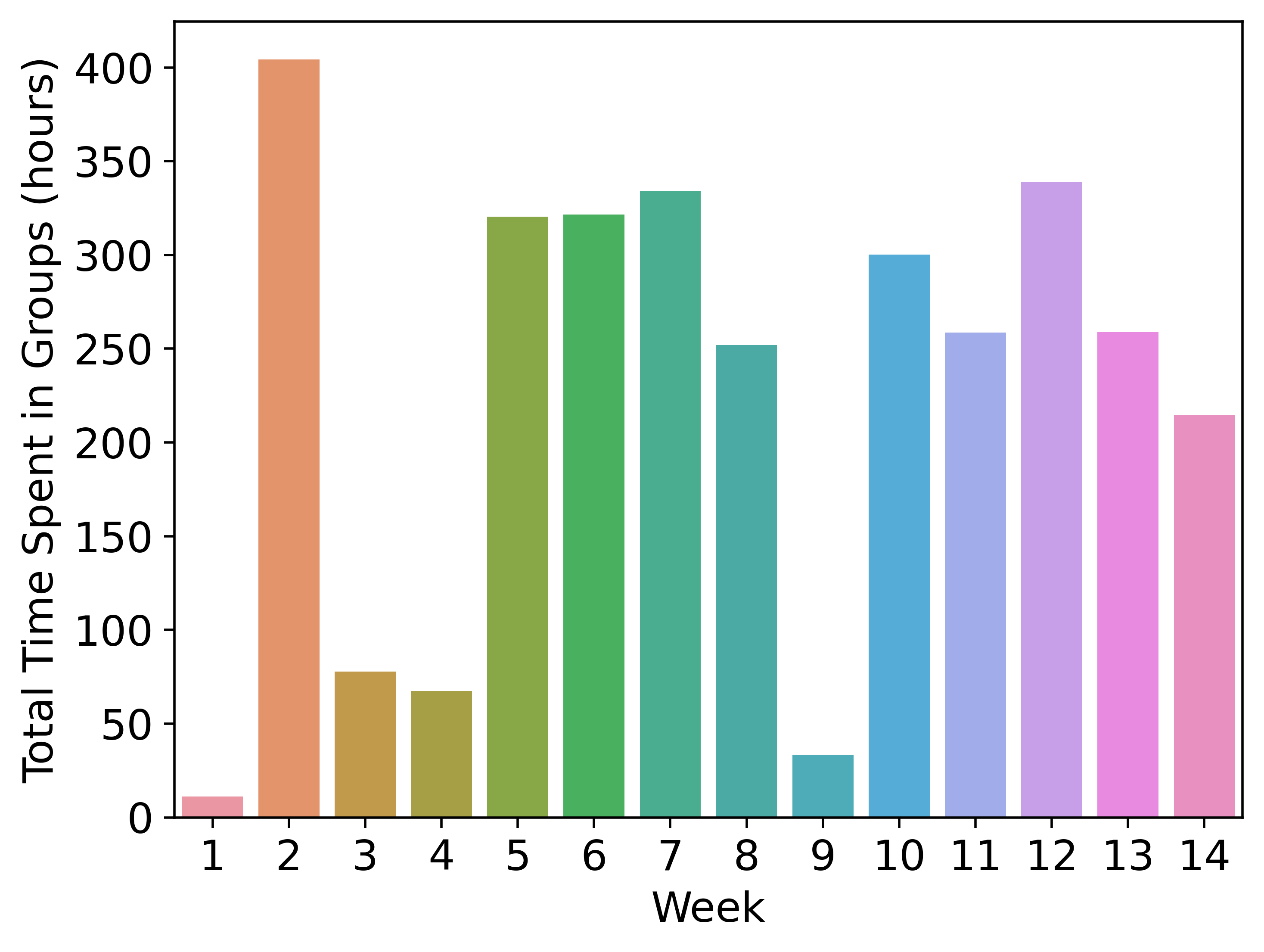}
    \caption{}
    \label{fig:cs_group_full}
    \end{subfigure}
    \begin{subfigure}[b]{0.6\textwidth}
    \includegraphics[width=\textwidth]{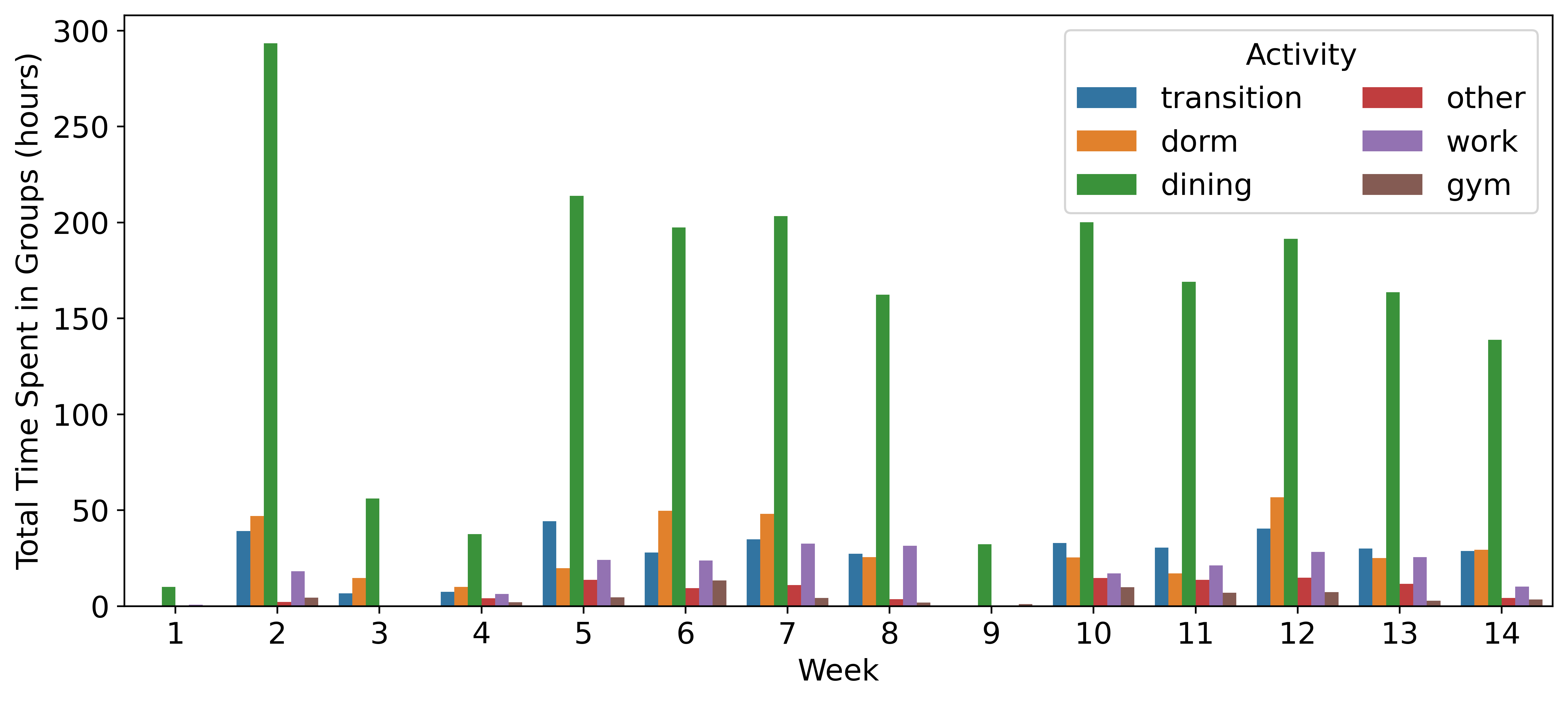}
    \caption{}
    \label{fig:cs_act_type}
    \end{subfigure}
    \caption{\minorreview{Group behavior over a semester: (a) total time spent weekly in groups; and (b) the respective activities.}}
    \label{fig:cs_group_fullact}
\end{figure}

\subsubsection{Time Spent and Activity Types in Groups}
The statistics in Figure \ref{fig:cs_group_full} show the variation of group sessions over 14 weeks, beginning one week before classes commenced and ending shortly before the final exams. We observe a sharp increase in group behavior corresponding to the first week of the semester (i.e., week 2). We expect this behavior to naturally reduce as students settle into their routines the following week but fluctuate as they move along with course requirements and midterms (i.e., weeks 6-8, weeks 8-11). In week 9, the significant reduction in group behavior is likely a result of Spring break, and this behavior fluctuates once again upon resuming the second half of the semester. 

Figure \ref{fig:cs_act_type} demonstrates the group activities that are most common on campus, along with how they change throughout the semester. From the plot, we can infer that students mostly gather over dining activities, consistent throughout the semester. Interestingly, we note that groups in dorms and work environments increase as we head toward midterm weeks 8 and 11.

Figure \ref{fig:cs2_time_hour} reveals the pattern of group behavior at an aggregated level on an hourly scale across the semester. The cumulative time for group sessions is significantly high throughout the morning, between 00:00 am to 5:00 am, likely because students are in their dorms and their devices are connected to the same WiFi AP.\footnote{Groups that repeatedly spend extended periods together in common locations such as dormitories or offices are not considered false positives as these groups are present for a common activity such as sleeping or working.} It should be emphasized that the variance for these group sessions is also high, suggesting many combinations of groups with varying levels of continuous WiFi connectivity while the device is not in use. Duration for these group sessions drops tremendously during the day as they leave their dormitory to fulfill their academic routines. Note that this drop in duration is also an artifact of the nature of passive sensing, in this case, WiFi, which often contains shorter continuous sessions after denoising and processing. \system can still successfully detect groups despite such noisy input.

\subsubsection{Activity Comparison Pre-Post COVID-19}
One of the earliest significant steps to curb the virus spread during the COVID-19 outbreak was for educational institutions to fully operate their academic schedules virtually, changing students' daily routines around campus. Over time, campuses took careful steps in reopening and implementing social restrictions at different phases. In this case, our analysis is to analyze the difference in group behaviors before and after COVID-19. 

\begin{figure}[t]
    \centering
    \begin{subfigure}[b]{0.35\textwidth}
    \includegraphics[width=\textwidth]{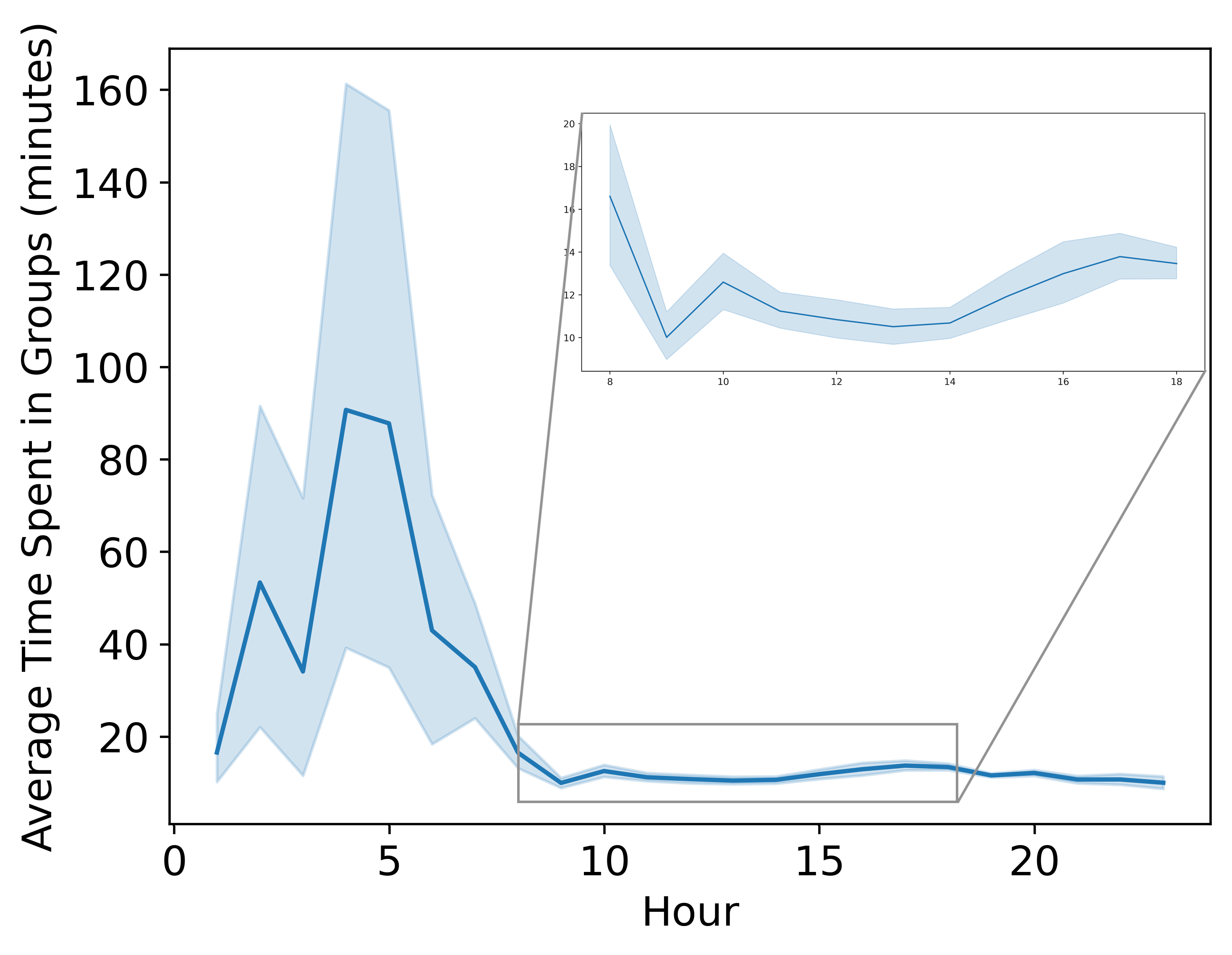}
    \caption{}
    \label{fig:cs2_time_hour}
    \end{subfigure}
    \begin{subfigure}[b]{0.6\textwidth}
    \includegraphics[width=\textwidth]{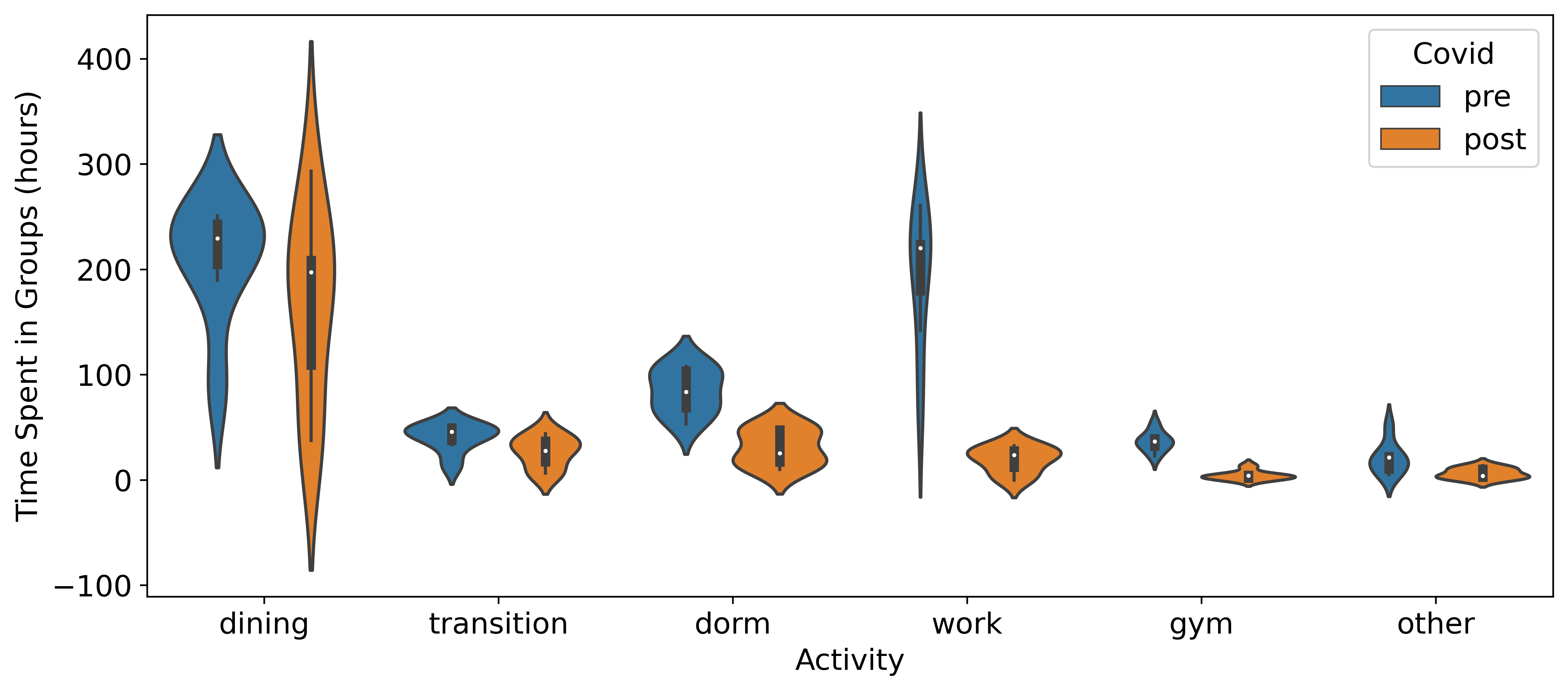}
    \caption{}
    \label{fig:cs2_prepost_covid}
    \end{subfigure}
    \caption{\minorreview{Group behavior over a semester: (a) average time spent in groups across a day; and (b) pre versus post COVID-19 comparison by activities.}}
    \label{fig:cs_group_fullact2}
\end{figure}

Figure \ref{fig:cs2_prepost_covid} compares two freshmen batches in 2019 and 2022. The violin chart illustrates the time spent weekly doing various group activities and its probability distribution. We observe a general trend of students spending less time in groups post-COVID. The time spent in groups pre-COVID has higher variability in dining and work activities. There is a higher distribution of time spent in different activities around the median post-COVID. Lesser time spent in dormitories and work sites implies fewer people live and work on campus, likely due to precautionary reasons.\\


\noindent \textbf{Key Takeaway:} \review{The results demonstrate how \system can be employed to analyze group behavior on a longitudinal scale. While the figures present intuitive information about student life on an aggregated scale, the ability to augment behavioral monitoring systems with information about social interactions can enable more robust psychological measures. This capability holds intriguing implications for behavioral, mental, and social health-related studies.}

\section{Discussion}
\label{sec:discussion}
Our study's objectives were to develop \system, \review{which quantitatively determines key characteristics of \wwww to distinguish group formations based on their purposes existing as short-term and long-term. Here we discuss the privacy and ethical implications of our system.}

\subsection{Strengthening Privacy Safeguards}
\review{\system's primary strategy to detect dynamic groups is utilizing mobility-typed data of any nature. It is worth noting that modern systems (e.g., WiFi networks, LBSNs) often rely on user authentication, which facilitates accessible data collection. Our user studies employed privacy safeguards of hashing device MAC and usernames, disallowing any reverse-engineering of the anonymization (see Section \ref{sec:study}, active users explicitly consented to be de-anonymized).} 

\review{Beyond the scope of this study, autonomous detection systems of digital users, such as ours, can raise privacy concerns. Further aggravating is the tendency for implicit consent via default opt-in, a widely used behavioral economics tool, requiring users to actively opt out should they choose differently \cite{nouwens2020dark}. 
Cumulatively, this often leads to the construction of detailed user profiles for targeted advertising and personalized content, with decisions about its collection and use made by powerful corporations \cite{cheney2017we}. Thus, we stress the significance of enabling users with greater control over their data for informed consent, aligning with the recommendations by Cheney-Lippold \cite{cheney2017we}. For example, users can opt into data sharing in particular vicinities, eliminating the inadvertant collection of sensitive data such as visit to a doctor's office, while still enabling group detection in many social settings.\footnote{\review{Such practices are automated in many LBSNs where location data is only shared across the network upon the user's explicit permission (e.g., Gowalla check-in dataset used in our evaluation).}} Nouwens et al. propose the usage of more durable design patterns for setting data preferences, effectively empowering users with the option to set controls upon initial authentication. Further abiding by Cheney-Lippold's recommendations on data security \cite{cheney2017we}, we recommend employing cryptographic techniques to standardize anonymized data collection in target environments.
As our case studies have demonstrated, \system's framework depends on the granularity of the data supplied. For example, in scenarios centering on large-scale behavioral monitoring, maintaining anonymity in data is sufficient to preserve user privacy while enabling meaningful aggregated analyses. Deanonymization is a second-step process necessary in applications requiring targeted user tracking. In such cases, while targeted users may be de-anonymized with their permission, the system can provide high-level insights on interactions with anonymized users relevant to the application at hand.}

\review{While cryptographic techniques stand to be well-accepted modes of data protection, we also consider potential improvements in modeling techniques that may enable model-dependent privacy protection. Specifically, we note that temporal discontinuity in mobility trajectories will produce shorter sessions in \system by design, which may not align perfectly with true group interactions. Our work continues to employ more sophisticated techniques to address these inaccuracies, such as adapting transformer-based models from natural language processing (NLP) to incorporate surrounding contexts and learn robust mobility representations that extend beyond spatiotemporal data \cite{vaswani2017attention}. Utilizing ML methods facilitates on-device and collaborative learning \cite{konevcny2016federated}, enhancing privacy by granting users greater control over their data and the ability to selectively retract specific portions of it while supporting community-driven group initiatives.}

\subsection{Application of Ethics}
\review{Consider our case study of analyzing longitudinal behavioral change among students on campus in Section \ref{sec:case-study-2} to separately analyze the behaviors of resident physicians in their workplaces \cite{adler2022burnout} as an example. Such a system would require individual stakeholders to trust the institution enormously. Thus, policy solutions are especially crucial as \system can be used without explicit user consent from its inherent mechanisms. A pressing concern requiring interdisciplinary research attention is how the system can provide actual benefits to target users without violating their privacy and feeling unfairly targeted or discriminated against \cite{benjamin2019assessing}. Monitoring technologies integrating community-based initiatives with personal benefits must be accompanied by appropriate policy frameworks protecting user rights, especially concerning health and well-being.}

\review{With \system accommodating different granularities of user permissions for personal health and productivity monitoring (e.g., monitoring employee burnout symptoms and work performance), it can facilitate usage across diverse ethical policies. For example, a multi-tiered crisis care infrastructure can be enabled among residents, supervisors, and educational mentors to foster communication among stakeholders and provide interventions. From a system's perspective, generalizing key group detection parameters from different mobility data sources ensures fair participation opportunities among users via commodity devices and services. Our evaluation of \system's robustness against ranging data characteristics in Section \ref{eval:gen} is guided by real-world inaccuracies in spatiotemporal signals. By providing residents the autonomy to share information types (e.g., location granularity, frequency, unique device identifier), \system readily adapts its grouping operation with a less precise one using anonymization techniques -- effectively storing the cohort's spatial, temporal, and social similarities at coarse granularity over time. Behavioral deviations from the cohort's norm (indicating a likelihood of burnout symptoms among unknown users) would reveal where and when these behaviors are observed, approximating what activities are typically being engaged. Without violating individual privacy, aggregated insights ensure consistent deployment of community crisis resources is within reach of potential target users. Aside from direct interventions, enabling a data-sharing use case to encourage training programs better manage residents' well-being \cite{adler2022burnout} should only be at the discretion of the individual, coupled with policies for data accountability.}

\review{In that order, \system's operationalization complies with the principles of the Belmont Report \cite{united1978belmont}. A framework capable of distinguishing real-world group formations via (1) common means to promote equal participation (\emph{justice}), (2) adaptation to changes in data granularity from users' autonomy (\emph{respect for persons}), and (3) maximizing possible benefits (\emph{beneficence}) across stakeholders in communities, can competently support practical and ethical applications of group dynamics.}

\section{Related Work}
\label{sec:related-work}
\review{The ubiquitous availability of sensing technologies has enabled researchers to comprehensively investigate user mobility and detecting groups at different granularity and modalities. The expansion of the ``digital nomad'' lifestyle motivated group modeling approaches to primarily seek location-based information \cite{lee2019social}. We summarize the key differences of these works in Table \ref{tab:summprior}.}

\begin{table}
    \centering
    \caption{\review{Comparison of prior work with ours. `Who' detects dynamically-changing small groups across time. `What' defines a detected activity. `When' analyzes whether group output was computed in time windows (periods) or at real-time. `Where' provides location information at varying granularity.}}
    \scalebox{0.75}{
    \begin{tabular}{l|ccc|c|cc|ccc} \toprule
    
    \multirow{3}{*}{Work(s)} & \multicolumn{3}{c|}{\textbf{Who}} & \textbf{What} & \multicolumn{2}{c|}{\textbf{When}} & \multicolumn{3}{c}{\textbf{Where}} \\ 
    
    & Dynamic & \makecell[c]{Small\\Group} & \makecell[c]{Long\\Term} & Activity & Periods & Real-Time & Coarse & Fine & Incomplete \\ \midrule
    
    Wearables \cite{olguin2008sensible, olguin2009capturing} & \cmark & \cmark & \xmark & \cmark & \xmark & \cmark & \xmark & \cmark & \xmark \\

    RFID Tags \cite{brown2014architecture} & \cmark & \cmark & \xmark & \xmark & \cmark & \xmark & \cmark & \xmark & \xmark \\ 

    Accelerometers \cite{gedik2018detecting} & \cmark & \cmark & \xmark & \xmark & \cmark & \xmark & \cmark & \xmark & \xmark \\ 

    BT Proximity \cite{feese2013sensing, yu2014grace, solmaz2017together} & \cmark & \cmark & \xmark & \xmark & \xmark & \cmark & \xmark & \xmark & \cmark \\ 
    
    BT Proxomity + Audio \cite{luo2013socialweaver, lee2013sociophone} & \cmark & \cmark & \xmark & \xmark & \xmark & \cmark & \xmark & \xmark & \cmark \\ 
    
    WiFi RSSI Signals \cite{hong2016socialprobe, shen2018snow, solmaz2020group, shen2020bag} & \xmark & \cmark & \xmark & \xmark & \cmark & \xmark & \xmark & \cmark & \cmark \\

    WiFi + Sensors \cite{kjaergaard2012detecting,du2017recognition} & \xmark & \cmark & \xmark & \xmark & \cmark & \xmark & \cmark & \cmark & \cmark \\
    
    WiFi Network Events \cite{sen2014grumon} & \xmark & \cmark & \xmark & \xmark & \cmark & \cmark & \cmark & \cmark & \cmark \\ 

    GPS \cite{wirz2011towards, yang2020extending} & \cmark & \cmark & \xmark & \xmark & \xmark & \cmark & \xmark & \cmark & \cmark \\
    
    LBSN Groups \cite{pham2013ebm, wang2014pgt, fan2017correlation, yang2019revisiting} & \xmark & \cmark & \xmark & \xmark & \cmark & \xmark & \xmark & \cmark & \cmark \\


    \rowcolor{lightgray}
    \system (ours) & \cmark & \cmark & \cmark & \cmark & \cmark & \cmark & \cmark & \cmark & \cmark \\ 
    \bottomrule
    \end{tabular}
    }
    \label{tab:summprior}
\end{table}

\subsection{Location-based social network (LBSN)} 
\review{LBSN derives group interactions based on user connections on content tied to locations. By gathering location data during user check-ins into the application, prior work has learned} the relationships between users where users' friendship networks serve as the ground truth for evaluation. Researchers have found that spatiotemporal co-occurrence strongly indicates social ties in LBSN datasets \cite{crandall2010inferring}. Many efforts went into understanding the relationship between mobility and strength of relationship \cite{li2008mining, eagle2009inferring, cranshaw2010bridging, pham2013ebm, wang2014pgt, fan2017correlation, yang2019revisiting}, while Xiao et al. utilized semantic information to estimate the similarity between people without geographical overlap \cite{xiao2014inferring}. More recent work introduced commitment and compatibility as properties that any social distance measure should follow to infer social connections \cite{pham2011towards}.

\subsection{Indoor Localization} 
\review{With humans spending most of their time indoors, researchers have increasingly investigated localization orientation by instrumenting IoT technologies within the environment or on-body wearables to derive group proximity. For example,} Brown et al. used lightweight RFID tags to detect face-to-face encounters  \cite{brown2014architecture}. Researchers also developed sociometric badges to measure interaction, physical proximity between users, and activity \cite{olguin2008sensible, olguin2009capturing}. More recently, Gedik and Hung used wearable accelerometers to identify conversational groups \cite{gedik2018detecting}. While these methods work well in accurately determining close contacts within an individual's social circle, a clear drawback is an insufficiency for scalability since these modalities require the dedicated installation of equipment, which is not feasible on a large scale. 

\subsection{Smartphone Localization} 
\review{The increasing reliance on smartphones by everyday users motivated researchers to model groups based on fixed and moving locations from the device or via its network infrastructures, such as Bluetooth, GPS, and WiFi. For example,} Feese et al. used a low-power ANT radio in smartphones to measure proximity to other firefighters during an event \cite{feese2013sensing}. Bluetooth (BT) signal strength is another modality for determining intentional and transitional groups \cite{yu2014grace, solmaz2017together}, including intersecting it with smartphone's microphone signals to detect conversational groups \cite{luo2013socialweaver, lee2013sociophone}. \review{Through spatiotemporal clustering techniques, prior work successfully used GPS to detect groups of pedestrians moving together \cite{wirz2011towards} and groups of vehicles on the road \cite{yang2020extending}.}

\review{Another modality showing promise in group detection is WiFi information, where mobility trajectories of users are estimated} from the access point (AP) connection events across time \cite{kaur2020joint} \review{and recursively approximating the distance between two users. Shen et al. proposed techniques using matrix factorization from the users' similarity matrix of raw WiFi logs with Bluetooth RSS data to detect groups \cite{shen2018snow,shen2020bag}. Group-In and SocialProbe, among others, are examples of group modeling using wireless RSSI \cite{solmaz2020group,hong2016socialprobe,sigg2014telepathic, gu2014wifi, gu2015paws}.} However, RSSI-based approaches have been criticized for being unreliable in crowded, urban spaces like malls and college campuses \cite{shen2020bag}. Since WiFi data is often incomplete and noisy, many similar approaches employ multimodal sensing to augment user mobility data with behavioral semantics \cite{kjaergaard2012detecting, sen2014grumon, du2016group, du2017recognition}. These efforts face similar challenges in scalability as sensor-based group detection methods.

\section{Conclusion}
\label{sec:conclusion}
This work proposed \system as a promising group detection framework that models and detects the \wwww \review{characteristics of group dynamics through a theoretical understanding of groups in small group research. \system incorporates the functional perspective of variability and repeatability to construct short-term and long-term groups through a distinctive view of \emph{spatial}, \emph{temporal}, and \emph{social} relationships.} These steps collectively allow the technique to more accurately determine users that colocate for a specific purpose and segregate those that colocate by chance. In this regard, our technique handles computational complexity to allow for scalability in detecting larger groups of users. Through \review{two} user \review{studies on three datasets}, our evaluation demonstrated \system yielding an average of 92\% overall accuracy, 96\% precision, and 94\% recall. Our results proved system robustness to different mobility-type datasets and performance improvements over prior group detection techniques. Finally, we \review{exhibited the practical relevancy of \system in} supporting two real-world applications, \review{further establishing the usability of the proposed system in applications requiring intentional group formations.}

\begin{acks}
We thank the anonymous reviewers for their valuable comments. This research was supported in part by NSF grants 2211302, 2211888, 2213636, 2105494, 1908536, Army Research Lab contract W911NF-17-2-0196, Adobe, and Ministry of Education, Singapore under its Academic Research Fund Tier 2 (Project ID: T2EP20220-0016). Any opinions, findings, and conclusions, or recommendations expressed in this material are those of the authors and do not necessarily reflect the views of the funding agencies.
\end{acks}

\bibliographystyle{ACM-Reference-Format}
\bibliography{ref}

\begin{appendices}
\section{Modeling \system}

\subsection{Fast Pairwise Co-Occurrence Detection}
\label{appx:fast-pairwise-occurrence}

The second module in \system requires the modeling of pairwise co-occurrences across all users. To address scalability concerns, we propose an approach that involves selectively comparing users and terminating the comparison process early when an overlap between pairs is detected, as outlined in Algorithm \ref{alg:module2}\footnote{Note that this method is equivalent to brute force comparison in the worst-case scenario where all users interact with each other. However, we work on the assumption that this is unlikely in most public settings.}.

We first sort the sessions by time for all users and subsetting sessions by location. These groups are filtered to be greater than size one (i.e., more than one session) and have more than two unique users. This step significantly narrows down the number of pairs that must be iterated. When iterating over groups, we use the sorted nature of the groups to identify when an overlapping session is found by comparing session entry and departure times. That is, for each user in each group, we iterate over sessions until the start time begins after the departure time of another user.

\begin{algorithm}
\caption{Fast Pairwise Co-Occurrence Detection}
\label{alg:module2}
\KwData{$S$}
\KwResult{$C$}
$S \gets $ sort($S$, by=e)\; 
S\_groups $\gets$ S.groupby(l)\;
S\_groups $\gets$ len(S\_groups) > 1 \& S\_groups[u].unique() > 2 \;
\For{$G$ in S\_groups}{
    \For{$i$ in $G$}{
        $j \gets i+1$\;
        found\_overlap $\gets$ False\;
        \While{!found\_overlap}{
              \eIf{$G_i[e] > G_j[d]$}{
                found\_overlap $\gets$ True\;
              }{
               $c_{ij}^t \gets \{G_j[e], min(G_i[d], G_j[d]), G_i[l], G_j[l], G_j[a]\}$
            }
            $j \gets j+1$
        }
    }
}
\end{algorithm}

\section{Prototype Implementation}

\subsection{Mobility Data Types}
\label{appx:data-collection}
Our current prototype supports two mobility data formats: WiFi mobility traces and location-based social network (LBSN) check-in data.

\subsubsection{WiFi mobility and location data:} \system can extract mobility and location traces of each user from logs collected by  WiFi access points in the form of syslog files. WiFi syslog data consists of AP connection events across time \cite{aruba}. As a user's device connects and disconnects to the APs, event logs are generated and aggregated in a centralized syslog server. Such data are routinely logged by commercial and consumer-grade access points. Further, many enterprises collect such information centrally using syslog servers for diagnosis and maintenance. All syslog messages are formatted as:
\begin{center}
\texttt{<date> <hh:mm:ss> <controller> <event\_ID> <severity> \\
<AP, MAC and IP addresses> <message text with BSSID and SSID>}
\end{center} 

The most relevant items in the log include \texttt{<date>}, \texttt{<hh:mm:ss>}, \texttt{event\_ID}, \texttt{AP} and \texttt{MAC}. \texttt{<date> <hh:mm:ss>} allow us to identify the precise timestamp of the event. The \texttt{event\_ID} contains the event type generated. We particularly focus on association, disassociation, reassociation, authentication, de-authentication, and drift events. When a user connects their device to the WiFi network, the device generates an association event. As the user moves with their device to another location, the device gets disassociated and re-associated with the nearest corresponding AP. Additionally, authentication events often contain messages with usernames and user roles (i.e., student or faculty/staff in a campus setting) that allow for the identification of username and device mapping. \texttt{AP} contains a unique identifier involving the building name, floor information, and room number. \texttt{MAC} or Media Access Control addresses contain the unique device ID. Note, the MAC addresses contained in our study dataset are hashed. We describe in Section \ref{sec:study}, participants consented to being detected through their hashed device MAC addresses.

The association and disassociation of MAC addresses to APs across buildings allow us to build a mobility trace of users. A fundamental assumption is that the users are connected to the same WiFi network. In order to remove inconsistencies and noise in the syslog data, the logs are grouped by MAC addresses, and each event is thoroughly analyzed to generate a sequence of association and disassociation logs. These logs are then interpolated as continuous mobility trajectories where missing entries are labeled with a "UNKN" location type.

For location types extraction, we classify college campus buildings into the following categories: administration, dining, health center, labs, landmark, library, parking, police, recreation, residential, school, student organizations, and others.
		
\subsubsection{LBSN check-in data:} \system can also extract location traces of a user using LSBN checkin data, which are  are a form of social networking data where the user connections and interactions revolve around content tied to locations. For instance, a user can share a message or content tied to a particular location, and other users can access this content when they arrive at the same location. Foursquare, Facebook Places, and Yelp are popular examples of LBSNs. LBSNs are formatted as:	
\begin{center}
\texttt{<user\_ID> <timestamp> <latitude> <longitude> <location\_ID>}
\end{center}

Embedded geo-locations in user check-ins or posts can be converted into mobility trajectories for each user. Since such data consists of location-based events at random intervals from users checking into the application, the data is often incomplete and not temporally continuous. However, \system is designed to support such data formats as well.

For location types extraction, we employ seven categories as in Liu et al. \cite{liu2014exploiting}: college campus buildings into the following categories: community, entertainment, food, nightlife, outdoors, shopping, and travel.
\end{appendices}

\end{document}